\documentstyle[editedvolume,numreferences]{crckapb} 

\input epsfig

\newcommand{\be}{\begin{equation}}
\newcommand{\ee}{\end{equation}}
\newcommand{\bea}{\begin{eqnarray}}
\newcommand{\eea}{\end{eqnarray}}
\newcommand{\nn}{\nonumber \\}
\newcommand{\p}[1]{(\ref{#1})}

\def\a{\alpha}\def\b{\beta}

\def\tr{\rm tr}
\def\Poinc{Poincar{\' e}}

\font\mybb=msbm10 at 10pt
\def\bb#1{\hbox{\mybb#1}}
\def\bZ {\bb{Z}}
\def\bR {\bb{R}}
\def\bE {\bb{E}}

\begin{opening}
\title{M-THEORY FROM ITS SUPERALGEBRA}
\subtitle{Carg{\`e}se Lectures 1997}
\author{P.K. TOWNSEND}
\institute{DAMTP, UNIVERSITY OF CAMBRIDGE\\
          SILVER ST., CAMBRIDGE CB3 9EW, U.K.}
\end{opening}
\runningtitle{M-THEORY FROM ITS SUPERALGEBRA}

\begin{document}
\begin{abstract}
These lectures explore what can be learnt about M-theory from its superalgebra.
The first three lectures introduce the `basic' branes of M-theory, and Type II
superstring theories, and show how the duality relations between them are
encoded in the respective spacetime superalgebras. The fourth lecture
introduces brane intersections and explains how they are encoded in the 
worldvolume superalgebras. 

\end{abstract}

\section{Lecture 1: M-branes and supersymmetry}

About twenty years ago, it was pointed out that D=11 is the maximum
dimension in which one can expect to find an interacting supersymmetric field
theory \cite{nahm}. Its existence, and uniqueness subject to standard
assumptions, was established with the construction of D=11 supergravity
\cite{CJS}. The linearized theory can be deduced from an analysis of the
standard D=11 super-\Poinc\ algebra and it will prove worthwhile to first
present a variant of this analysis. Since the commutators involving Lorentz
generators are determined by the Lorentz indices it is convenient to concentrate
on the subalgebra of supertranslations spanned by the 11-momentum
$P_M$ and the 32-component Majorana spinor charge $Q_\a$. Translation invariance
implies that
$P$ and $Q$ commute, leaving the anticommutator
\be\label{onea}
\{ Q_\alpha,Q_\beta\} = (C\Gamma^M)_{\alpha\beta} P_M
\ee
as the only non-trivial relation, where $\Gamma^M$ are the Dirac matrices and 
$C$ is the (real, antisymmetric) charge conjugation matrix. There are actually
two inequivalent representations of the Dirac matrices. They differ according to
whether the product of all 11 of them is $1$ or $-1$. The choice is arbitrary
and we shall suppose that
\be\label{oneb}
\Gamma^0\Gamma^1\cdots \Gamma^9\Gamma^\natural =1\; .
\ee
The symbol $\natural$ is used here to denote the number 10. It is convenient to
have a single symbol for this number because we shall be using a convention in
which $\Gamma^{10}$ will mean the product $\Gamma^1\Gamma^0$. More generally,
the antisymmetrized product of $n$ Dirac matrices will be written as
$\Gamma^{M_1\dots M_n}$. Thus (\ref{oneb}) is equivalent to $\Gamma^\natural =
\Gamma^{0\cdots 9}$ (also known as $\Gamma_{11}$ in the context of D=10 IIA
supergravity).

Let us suppose that there exists some quantum theory that realizes the algebra
(\ref{onea}) as an asymptotic symmetry, e.g. of the S-matrix. The asymptotic
metric is assumed to be the D=11 Minkowski metric and the vacuum is assumed to 
be annihilated by all supersymmetry charges. Consider  a state
preserving some non-zero fraction $\nu$ of the supersymmetry of the vacuum.
This state will be annihilated by some combination of supersymmetry charges and
the expectation value of $\{Q,Q\}$ will be a real symmetric positive
semi-definite matrix with $32\times\nu$ zero eigenvalues. In particular, its
determinant will vanish. From (\ref{onea}) we see that this will happen when
\be\label{onec}
0=\det \Gamma\cdot P = (P^2)^{16}\; .
\ee
Thus, given the algebra (\ref{onea}), we deduce that the only states preserving
a non-zero fraction of the supersymmetry, apart from the vacuum, are those
for which the 11-momentum is null. To determine the fraction $\nu$ we note that
if $P$ is null we can choose a frame in which (the metric convention is
`mostly plus' and $P^0$ is the energy)
\be\label{oned}
P_M = {1\over2}(-1;\pm 1,0,\dots,0)\; .
\ee
We may also choose the Majorana (real) representation of the Dirac matrices for
which $C=\Gamma^0$. The algebra (\ref{onea}) is then
\be\label{onee}
\{ Q_\alpha,Q_\beta\} = {1\over 2}(1\mp \Gamma_{01})_{\alpha\beta}
\ee
where $\Gamma_{01}=\Gamma_0\Gamma_1$, and similarly for other products of Dirac
matrices. Clearly, eigenspinors of $\{Q,Q\}$ with zero eigenvalue satisfy
\be\label{onef}
\Gamma_{01}\epsilon =\pm\epsilon\; .
\ee
Since $\Gamma_{01}$ squares to the identity its eigenvalues are $\pm1$, and
since it is also traceless precisely half are $+1$ and half $-1$. Thus, the 
space of solutions to (\ref{onef}) is 16-dimensional and we therefore deduce that
$\nu=1/2$. {\it There are no other possibilities allowed by the supersymmetry
algebra (\ref{onea})}.

It is natural to associate the $\nu=1/2$ states for which $P^2=0$ with massless
particles. Classically, and in the absence of interactions, a massless
particle in a D=11 Minkowski vacuum is described by the action 
\be\label{oneg}
S[X; e]= \int\! d\tau\, e^{-1} g 
\ee
where the `einbein' $e(\tau)$ is an independent worldline scalar density and
$g$ is the metric on the worldline induced from spacetime, i.e. $g= \dot X^2$.
In the D=11 Minkowski vacuum, this action is invariant under translations of $X$.
The corresponding Noether charge is the null 11-momentum $P$. To incorporate
supersymmetry we must replace the translation-invariant differential $dX$ by the
supertranslation-invariant differential $\omega=dX-i\bar\theta\Gamma d\theta$, so
that $g= \omega^2$. The worldline `fields' of the resulting `superparticle'
action are therefore the bosons $X^M(\tau)$ and the fermions $\theta_\a (\tau)$.
The fermion `fields' $\theta_\a$ transform inhomogeneously under supersymmetry
($\theta\rightarrow \theta + \epsilon$) and can be interpreted as 
Nambu-Goldstone (NG) variables for broken supersymmetry. This would appear to
imply that all supersymmetry is broken but the massless superparticle action has
a fermionic gauge invariance, known as `$\kappa$-symmetry', that allows half the
fermions to be `gauged away'. There are therefore only 16 NG variables,
corresponding to $\nu=1/2$. On the one hand this symmetry appears miraculous
since we made no attempt to build it into the construction. On the other hand
there would have been a contradiction with our previous analysis had it not been
present. In any case, only half of the 32 components of $\theta(\tau)$ are
physical. These 16 physical fermion fields may be split into 8 creation
operators and 8 annihilation operators. Quantization of the massless
superparticle then leads to linearized field equations for a supermultiplet of
D=11 supersymmetry with $2^8$ components, of which 128 are bosons and 128
fermions. Detailed investigation reveals that the Lorentz representations are
those of the graviton supermultiplet of D=11 supergravity, i.e. the graviton,
gravitino, and 3-form gauge potential $A$. In
other words, quantization of the massless D=11 superparticle yields the
linearized field equations of D=11 supergravity. 

By including interactions we arrive at the full non-linear field equations of
D=11 supergravity. We may now ask whether these equations admit solutions
preserving some fraction of the supersymmetry. The procedure for doing this is
to seek D=11 spacetimes admitting Killing spinor fields, i.e. spinors statisfying
the first order differential equation
\be\label{oneh}
\delta_\epsilon \psi \equiv {\cal D} \epsilon =0\; ,
\ee
where $\psi$ is the one-form gravitino field and $\delta_\epsilon\psi$ its
supersymmetry variation with spinor parameter $\epsilon(x)$. This variation
defines a covariant exterior derivative ${\cal D}$. In principle, this operator
includes terms involving contractions of Dirac matrices with the 4-form field
strength $F=dA$. These terms will eventually prove to be of crucial importance
but they play no role in the present discussion. Setting $F=0$, the general
11-metric admitting Killing spinors may be shown to take the `M-wave' form
\cite{hull}
\be\label{onei}
ds^2 = du dv + K({\bf x},u) du^2 + d{\bf x}\cdot d{\bf x}\; ,
\ee
where ${\bf x}$ are cartesian coordinates for $\bE^9$ and $K$ is an arbitrary
function of $u$ but is harmonic on $\bE^9$. This metric is
asymptotically flat provided that $K\rightarrow 0$ as $|{\bf x}|\rightarrow
\infty$. In this case there is an asymptotically flat region in which $P$ is
defined by an ADM-type formula, and one can verify that $P^2=0$. If the
above metric is used to compute ${\cal D}$ one finds that the condition
(\ref{oneh}) reduces to $\Gamma_v\epsilon=0$, which (setting $v=t+x^1$) is
equivalent to (\ref{onef}). Actually, the spinor $\epsilon$ here is a field
but it is assumed to be constant at infinity and this constant
spinor may be identified with a constant zero-eigenvalue eigenspinor of
$\{Q,Q\}$. 

Given a solution of D=11 supergravity one may ask for the effective action
determining the dynamics of small fluctuations about it. This dynamics is
governed by the Nambu-Goldstone variables associated with the symmetries broken
by the solution and these variables are fields on the orbits of Killing vector
fields of the background. The M-wave breaks translational invariance in the $(u,
{\bf x})$ directions (generically) and 1/2 of the supersymmetry. It has
$\partial/\partial v$ as its only Killing vector field. We conclude that the
effective action for small fluctuations is a functional of the `fields'
$\big(u(v), {\bf x}(v); \lambda(v) \big)$, where $\lambda$ represents the 16
NG fermions of 1/2-broken supersymmetry. Symmetry considerations now imply that
the action must be the light-cone gauge-fixed version of the massless
superparticle. We have now arrived back at our starting point in the analysis
of $\nu=1/2$ realizations of the standard D=11 superymmetry algebra.

About ten years ago it was realized that it is possible to couple D=11
supergravity to a closed supermembrane \cite{BST}. The membrane action has the
form
\be\label{onej}
S= -T_2 \int vol(g) +  Q_2 \int\! {\cal A}
\ee 
where $vol(g)$ is the worldvolume density in the induced metric $g$ and
${\cal A}$ is the pullback to the worldvolume of the 3-form potential $A$. 
The constant $T_2$ is the surface tension, while $Q_2$ is a membrane charge. The
supermembrane action is formally the same, but $vol(g)$ and ${\cal A}$ are
induced from superspace. Actually, the supermembrane action of \cite{BST} is the
special case of (\ref{onej}) in which $Q_2=T_2$. It turns out that precisely in
this case the action is $\kappa$-symmetric, i.e. it has a fermionic gauge
invariance that allows half of the worldvolume fermion fields to be `gauged
away'. More precisely, the supermembrane action is $\kappa$-symmetric if the
background satisfies the field equations of D=11 supergravity. This allows, in
particular, a D=11 vacuum background, in which the worldvolume fermions can be
interpreted as the NG variables associated with the breaking by the membrane of
1/2 the  supersymmetry of the vacuum. This presents us with a paradox because we
have just shown that the  D=11 supersymmetry algebra (\ref{onea}) allows only
one object to have this property, a massless particle. The resolution of this
paradox is that the supersymmetry algebra is modified in the presence of a
membrane. In the D=11 Minkowski vacuum the supermembrane  action is
supertranslation invariant and there therefore exist conserved Noether charges
$P_M$ and $Q_\a$. However, these charges do not obey the algebra (\ref{onea}).
The reason for this is that, unlike the massless superparticle action, the
supermembrane action is not {\it manifestly} supersymmetric: the 
coupling to the 3-form potential $A$ does not vanish in the D=11 Minkowski vacuum
but instead reduces to a Wess-Zumino term for the super-\Poinc\ group. This term
(which is defined up to the addition of an exact 3-form) can be chosen to be
translation invariant but it then changes under a supersymmetry transformation
by the addition of an exact 3-form and this leads to a modification of the
algebra of the supertranslation Noether charges
\cite{jerome}. The modified algebra is   
\be\label{onek}
\{ Q_\alpha,Q_\beta\} = (C\Gamma^M)_{\alpha\beta} P_M + {1\over2}
(C\Gamma_{MN})_{\alpha\beta} Z^{MN}\; ,
\ee
where
\be\label{onel}
Z^{MN}= Q_2\int dX^M \wedge dX^N\; ,
\ee
and the integral is taken over the 2-cycle occupied by the membrane in
spacetime. The 2-form charge $Z$ is not central in the super-Poincar{\'e} algebra
because it does not commute with Lorentz transformations, but this is to be
expected of a charge carried by an extended object\footnote{The fact that
p-form charges are carried by p-branes was the main result of \cite{jerome}
but the charges themselves were considered previously in the context of
the group manifold approach to supergravity \cite{fre}, on mathematical grounds
\cite{vans}, in the context of super Yang-Mills theory \cite{vanNic,Zizzi}, and
in some other `p-brane-inspired' generalizations of the supersymmetry algebra
\cite{bergs,sezgin}. A related modification of the worldvolume
supersymmetry algebra of gauge-fixed super $p$-branes was found in
\cite{HLP}.}. The charge $Q_2$ is thus analogous to the string winding number in
string theory; it vanishes unless the two-cycle is non-contractible. The case of
an infinite planar membrane can be dealt with by considering it as a limit of
one wrapped on a large torus. In this limit $P^0$ and some components of $Z$ are
infinite but the tension $T_2$ and the charge $Q_2$ remain finite.  

Essentially, the supermembrane is an exotic form of matter. As for any other
form of matter, a membrane will produce long-range gravitational and other
fields, allowing the tension $T_2$ and charge $Q_2$ to be detected as surface
integrals at infinity. In the case of an infinite planar membrane,
`infinity' should be interpreted to mean `transverse spatial infinity', i.e. a
large distance limit in non-parallel directions. Transverse spatial infinity is
topologically $\bR^2\times S^7$, but the asymptotic translational invariance in
directions parallel to the membrane allows us to reduce integrals for total
charges to integrals over $S^7$ for charge densities. The energy density is the
surface tension $T_2$,  which is given by a modification of the usual ADM
formula \cite{Lu}. The membrane charge density is given by
\be\label{onem}
Q_2 = \Omega_7^{-1}\int_{S^7} \star F
\ee 
where $F=dA$ is the 4-form field strength and $\Omega_7$ the volume of the unit
7-sphere. The integral should be evaluated at infinity because, in contrast to
electrodynamics, the field equation for $A$ is not $d\star F=0$, but rather 
\be\label{onen}
d[\star F + F\wedge F]=0\; ,
\ee
where the second term is due to the Chern-Simons (CS) term in the action. Thus,
$\star F$ is not necessarily a closed 7-form. We shall consider the implications
of this below, but $d\star F$ will vanish asymptotically in the circumstances
described above. 

From \p{onel} we see that a membrane in the 12 plane is associated with
non-zero $Z_{12}$. Let us again choose the Majorana representation of the Dirac
matrices in which $C=\Gamma^0$. Then, for a static membrane, the algebra 
\p{onea} becomes
\be\label{oneo}
\{ Q,Q\} = P^0 + \Gamma^{012}Z_{12}\; .
\ee
Now, $Q$ is real in the Majorana representation, so the left hand side is
manifestly positive. Since the sign of $Z_{12}$ can be flipped by replacing the
membrane by an anti-membrane we must have $P^0\ge0$. If $P^0=0$ we have the
vacuum. Otherwise $P^0 >0$ and we derive the Witten-Olive-type bound $P^0 \ge
|Z_{12}|$, which is equivalent to
\be\label{onep}
T_2 \ge |Q_2|\; .
\ee
A stable membrane is expected to saturate this bound, so the case in which the
bound is saturated, i.e. $T_2=|Q_2|$, is of particular importance. In this
case the anticommutator \p{oneo}
becomes
\be\label{oneq}
\{ Q,Q\} = P^0 [1 \mp \Gamma^{012}]\; .
\ee
Spinors $\epsilon$ satisfying
\be\label{oner}
\Gamma^{012}\epsilon =\pm \epsilon
\ee
are eigenspinors of $\{Q,Q\}$ with zero eigenvalue. Since $\Gamma^{012}$
squares to the identity, and is traceless, the dimension of the zero-eigenvalue
eigenspace of $\{Q,Q\}$ is 16. In other words, a membrane saturating the bound
\p{onep} preserves 1/2 the supersymmetry of the vacuum \cite{jerome}. This fact
is directly related to the $\kappa$-symmetry of the supermembrane action.

The bound (\ref{onep}) can also be derived \cite{ght}, subject to standard
assumptions, via a modification of the Gibbons-Hull bound on the mass of charged
black holes in General Relativity\footnote{Actually, in the conventions used in
\cite{ght}, the bound is saturated when $T_2=(1/2)|Q_2|$, and there is a similar
factor of 1/2 for the fivebrane to be discussed below. To retain these factors
would suggest a level of attention to the consistency of conventions that has
not been attempted here. Various factors in formulas appearing in these lectures
have therefore been set to one on the grounds that this is the correct factor
for some convention.}. In this approach, the solutions saturating the bound are,
as in the M-wave case, those admitting Killing spinor fields satisfying
(\ref{oneh}), but now one must allow for a 4-form field strength $F$ consistent
with the asymptotic form required by a non-zero membrane charge 
$Q_2$. In this way one obtains the following 1/2 supersymmetric membrane
solution of D=11 supergravity \cite{ds}:
\bea\label{ones}
ds^2 &=& H^{-2/3}ds^2(\bE^{2,1}) + H^{1/3}ds^2(\bE^8) \nn
F &=& vol(\bE^{2,1}) \wedge dH^{-1}
\eea
where $H({\bf x})$ is a harmonic function on $\bE^8$. The choice
\be\label{onet}
H = 1 + \sum_{s=1}^n {|q_s|\over |{\bf x} -{\bf x}_s|^6}
\ee
leads to a metric that is `transverse-asymptotically' flat and which
can be interpreted as $n$ parallel membranes with charges $q_s$ at `centres'
${\bf x}= {\bf x}_s$, ($s=1,\dots,n$). The singularities at the centres of the
metric are actually degenerate Killing horizons. The maximal analytic extension
is analogous to the extreme Reissner-Nordstom (RN) multi black hole solution of
Maxwell-Einstein theory in D=4; there is a singular timelike membrane source
behind each horizon \cite{dgt}. The existence of multi-membrane solutions
indicates that the static force between parallel membranes cancels because of a
balance between the attraction due to gravity and the electrostatic-type
repulsion due to the 3-form potential. This balance is possible only when
$T_2=|Q_2|$. Again, there is a close analogy here to extreme RN black holes.

The two-form central charge is not the only possible central extension of the
D=11 supertranslation algebra. It is also possible to include a five-form 
charge \cite{fre,vans}. Given that the two-form extension is associated with a
membrane it might seem obvious that the five-form is associated with a
fivebrane, but it actually took almost another ten years for this connection to
be made. To see how a fivebrane might appear, consider the surface integral 
\be\label{oneu}
Q_5 = \Omega_4^{-1} \int_{S^4} F\; ,
\ee
where $\Omega_4$ is the volume of the unit 4-sphere.
Because of the Bianchi identity $dF=0$, this integral is homotopy invariant: if
it is non-zero the 4-sphere must surround a magnetically-charged topological
defect of the 3-form potential $A$. This defect must be 5-dimensional, and
the 4-sphere can be taken to be a sphere of fixed radius in a 5-dimensional
space transverse to the 6-dimensional worldvolume of a fivebrane defect. This is 
just an application to D=11 and $p=2$ of the well-known fact that
the magnetic dual of a p-brane in a D-dimensional spacetime is a
$(D-p-4)$-brane.  Assuming boundary conditions appropriate to an infinite planar
fivebrane one can derive the bound $T_5 \ge |Q_5|$ on the fivebrane tension. The
configuration saturating the bound is \cite{guven}
\bea\label{onev}
ds^2 &=&  H^{-1/3} ds^2(\bE^{(5,1)}) + H^{2/3} ds^2(\bE^5) \nn
F &=&  *dH
\eea
where $*$ is the Hodge dual on $\bE^5$ and $H({\bf x})$ is a harmonic function
on $\bE^5$. The choice
\be\label{onew}
H= 1 + \sum_{s=1}^n {|p_s|\over |{\bf x} -{\bf x}_s|}
\ee
can be interpreted as $n$ parallel fivebranes with charges $p_s$ at positions
${\bf x} ={\bf x}_s$, ($s=1,\dots,n$) in $\bE^5$. Like the membrane solution of
D=11 supergravity, this fivebrane solution admits 16 Killing spinor fields
and so preserves half the supersymmetry. The singularities of $H$
are again just Killing horizons of the metric, but in this case there is no
singularity behind the horizon; the maximal analytic extension is geodesically
complete \cite{ght}. 

The fact that $dF=0$ means that the 4-sphere surrounding the fivebrane can be
deformed at will without changing the value of the charge $Q_5$. We do not even
have to worry about passing through singularities of $F$ because, as just
mentioned, there are none. One consequence of this is that a fivebrane carrying
non-zero $Q_5$ charge must be closed, i.e. it cannot have a boundary. If there
were a boundary then the 4-sphere of the $Q_5$ integral could be pushed past it
and shrunk to a point, so $Q_5$ would vanish, contrary to hypothesis. A
similar conclusion would hold for the membrane if the equation of motion for
the 3-form potential were $d\star F=0$, but as this is not the equation of
motion the conclusion is modified. A closer analysis \cite{surgery} shows that
the membrane may have a boundary on a fivebrane\footnote{A conclusion
originally arrived at in \cite{strom,DfromM}, following the discovery that Type
II superstrings can end on D-branes \cite{RRpolch}.}. We pass over the details
here as we shall eventually recover the result from a different approach.

By analogy with the membrane, one would expect the fivebrane charge $Q_5$ to be
the magnitude of a 5-form charge in the D=11 supertranslation algebra. In
other words, taking both the membrane and the fivebrane into account we   
that \p{onek} should be replaced by \cite{democracy}
\be\label{onex}
\{ Q_\alpha,Q_\beta\} = (\Gamma^MC)_{\alpha\beta} P_M + {1\over2}
(\Gamma_{MN}C)_{\alpha\beta} Z^{MN} +
{1\over 5!} (\Gamma_{MNPQR}C)_{\alpha\beta} Y^{MNPQR} 
\ee
where $Y$ is the 5-form charge.
One indirect argument for its presence is that the
fivebrane solution could not otherwise be half-supersymmetric. To see that this
{\it is} possible in the presence of a 5-form charge, let us associate a static
fivebrane in the 12345 5-plane with non-zero $Y_{12345}$, and define $q_5=
Y_{12345}/P^0$. The $\{Q,Q\}$ anticommutator is then
\be\label{oney}
\{Q,Q\} = P^0\big[1 + \Gamma^{012345}q_5\big]\, .
\ee
As before, we deduce that $|q_5|\le 1$. We may identify $|q_5|$ with
the ratio of the fivebrane's charge $Q_5$ to its tension $T_5$, so the bound is
equivalent to $T_5\ge |Q_5|$. When this bound is saturated, half of the
eigenvalues of $\{Q,Q\}$ vanish so that half the supersymmetry is preserved. 

If the fivebrane is to be associated with a 5-form charge in the supersymmetry 
algebra then its worldvolume action must contain a coupling to a 6-form
potential ${\cal B}$ induced from a 6-form $B$ on superspace. It is natural to
identify $B$ as the magnetic potential dual to $A$, but the presence of the CS
term in the D=11 supergravity action seems to prevent the dualization of $A$. 
However, the $A$ field equation allows us to write 
\be\label{onez}
[\star F + F\wedge F] = d B\; ,
\ee
and there is a superfield version of this \cite{Lechner}. Thus, the
required superspace 6-form potential exists for on-shell supergravity
backgrounds\footnote{Both $A$ and $B$ appear off-shell in a new formulation of
D=11 supergravity \cite{berk}.}. The on-shell restriction is no disadvantage
since this is in any case an expected consequence of $\kappa$-symmetry.

One might suppose that the superfivebrane action is just a
higher-dimensional generalization of the supermembrane action, but this is not
so. In both cases the worldvolume field theory must, after (partial) gauge
fixing, have 16 (linearly realized) supersymmetries (counting each component of
the supercharge separately). Massless supermultiplets of such supersymmetry
algebras have 8 boson degrees of freedom, per worldvolume point (16 in phase
space). In the membrane case the worldvolume fields are the maps from the
worldvolume to superspace so that on fixing the 3 worldvolume diffeomorphisms
we are left with 8 physical scalar fields describing transverse fluctations of
the membrane. For the fivebrane a similar count yields 5 scalars describing
fluctations transverse to the fivebrane. We therefore need 3 additional boson
degrees of freedom, which must be provided by other boson field(s) needed to
complete a supermultiplet of N=2 D=6 supersymmetry. The only possibility is the 
antisymmetric tensor multiplet of chiral (2,0) D=6 supersymmetry
\cite{gibtown,kaplan} (originally identified as the one containing the
worldvolume fields of the IIA fivebrane \cite{CHS}\footnote{In that case only
four of the five scalars were interpreted as NG variables representing
transverse fluctuations, but the presence of the fifth scalar suggests an
11-dimensional interpretation. In fact, the fivebrane solution of D=10 IIA
supergravity {\it is} the fivebrane solution of D=11 supergravity; one has only
to reinterpret the fields.}). As the terminology suggests, the (2,0)
antisymmetric tensor multiplet has a 2-form gauge potential
$U$. Its field strength $H=dU$ is self-dual. This makes the construction 
of an action
difficult, even at the linearized level. Some of these difficulties are
intrinsic and lead to unavoidable complications in the quantum theory
\cite{Witfive}. One could ask only for field equations, and these were presented
in \cite{howe} as superfield constraints on the extrinsic supergeometry. One
could also abandon manifest six-dimensional Lorentz covariance, and an action of
this type was given in \cite{agan}. However, it is possible to find
a manifestly Lorentz invariant superfivebrane action \cite{pasti}, at the cost 
of having to include a factor of $1/(\partial a)^2$ where $a$ is an additional
worldvolume scalar gauge field (the `PST field'). These three formulations have
been shown to be classically equivalent \cite{HSW,pastib}, and the Noether
charges of the Lorentz covariant action have been shown to obey the algebra
(\ref{onex}), with both 5-form and  2-form charges \cite{dmitri}. 

The action of \cite{pasti} provides a simple route to the hamiltonian
formulation, some aspects of which will be explained here for later use. This
formulation is especially simple in a vacuum background. Omitting fermions, the
phase-space Lagrangian density is \cite{bergtown}
\bea\label{phasea}
{\cal L} &=& P\cdot \dot X + \Pi^{ab} \dot U_{ab} + \lambda_a \partial_b \Pi^{ab}
- s^a\big( P\cdot \partial_a X -V_a \big) \nn
&& +\, \sigma_{ab}(\Pi^{ab} + {1\over4}\tilde {\cal H}^{ab}) - {1\over 2}v 
\big[ (P- g^{ab}V_a\partial_b X)^2 + \det (g+ \tilde H) \big]
\eea
where $g$ is here the `worldspace' 5-metric of the fivebrane and
\bea\label{phaseb}
\tilde {\cal H}^{ab} &=& {1\over6} \varepsilon^{abcde} H_{cde}\nn
\tilde H_{ab} &=& {1\over \sqrt{\det g}} g_{ac} g_{ad} \tilde {\cal H}^{cd}\nn
V_f &=& {1\over 24} \varepsilon^{abcde} H_{abc} H_{def} 
\eea
Note that the Gauss law constraint on the electric 2-form
$\Pi$ (imposed by the Lagrange multiplier $\lambda_a$) becomes equivalent
to the Bianchi identity $dH=0$ on using the constraint imposed by the Lagrange
multiplier $\sigma_{ab}$. This is how the self-duality of the worldvolume
3-form $H$ is incorporated in the phase-space action, which depends only the
worldspace components of $H$. The Lagrange multipliers $s^a$ and $v$ impose the
worldspace diffeomorphism and hamiltonian constraints, respectively.

At this point it may seem that a new term in the D=11 supersymmetry algebra is
found every ten years and that it might therefore be prudent to wait another ten
years to see what happens. While there will undoubtedly be many advances in our
understanding of M-theory over the next ten years, the addition of another
term to the D=11 supersymmetry algebra is not likely to be one of them. The total
number of entries of the real symmetric $32\times 32$ matrix $\{ Q,Q\}$ is $528$.
This is the same as the total number of components of $P$, $Z$ and $Y$. To
see that this is no accident we first note that the matrices $C\Gamma^{M_1\dots
M_n}$ are either symmetric or antisymmetric, depending on the value of $n$.
Because of \p{oneb}, we need consider only $n\le5$ and of these only the
$n=1,2,5$ matrices are symmetric. Thus, (\ref{onex}) is the most general D=11
super-translation algebra with one D=11 spinor charge. The matrix $\{Q,Q\}$ may,
initially, be considered as an element in the Lie algebra of the group
$Sp(32;\bR)$. This has an $SO(1,10)$ subgroup with respect to which the
528-dimensional adjoint representation of $Sp(32;\bR)$ has the following
decomposition
\be\label{lasta}
{\bf 528} \rightarrow {\bf 11} \oplus {\bf 55} \oplus {\bf 462}\, .
\ee
This is just the decomposition provided by the charges $P$, $Z$ and $Y$.
In fact, because the bosonic charges are all assumed to be abelian, the algebra
spanned by $(Q;P,Y,Z)$, is a contraction of $OSp(1|32;\bR)$. 

The algebra (\ref{onex}) was called the `M-theory algebra' in \cite{fourM}
because it encodes many of the important features of M-theory. The most
significant is that it shows that, in addition to M-waves, both membranes and
fivebranes, henceforth to be called M-2-branes and M-5-branes, can also preserve 
1/2 supersymmetry. In this sense, and also because they are related by dualities
(in a way to be explained below) the M-Wave, M-2-brane and M-5-brane should be
considered on a similar footing. However, these are not the only ingredients. In
compactified spacetimes there are additional possibilities, as we now discuss.

\section{Lecture 2: More branes from M-theory}

Membranes and fivebranes are associated with the {\it spatial} components of the
charges $Z$ and $Y$ in the M-theory superalgebra. Having introduced these
charges we should now consider what might be the significance of their time
components. This is most straightforward for the 5-form charge $Y$, so we
consider it first. Let us suppose that all bosonic charges other than
$P^0$ and $Y_{0789\natural}$ vanish, and set $\tilde q_5 = Y_{0789\natural}/P^0$.
Then
\be\label{twoa}
\{ Q_\alpha,Q_\beta\} = P^0\big[1-\Gamma_{789\natural}\; \tilde q_5 \big]\, .
\ee
Again, $\Gamma_{789\natural}$ squares to the identity and has zero trace, and
therefore has eigenvalues $\pm1$. Positivity implies the bound
$|\tilde q_5|\le1$ and configurations which saturate the bound preserve
1/2 supersymmetry and are associated with the constraint
\be\label{twob}
\Gamma^{789\natural}\epsilon =\pm \epsilon\, .
\ee
This constraint is not obviously associated with a brane, but it
is equivalent to
\be\label{twoc}
\Gamma^{0123456}\epsilon =\pm\epsilon
\ee
which suggests a 6-brane. There is no 7-form gauge potential of D=11 
supergravity to couple to a 6-brane but if we compactify in the
$\natural\,$th direction on $S^1$ we obtain a Kaluza-Klein (KK) vector potential
$C$ in D=10, with 6-form dual. Thus, an object that is magnetically charged with
respect to $C$ is a 6-brane. This 6-brane has an M-theory interpretation
\cite{revisited} as a KK monopole \cite{sorkin,KKmon}. 

To see this in more detail we need the KK ansatz relating the bosonic fields of
D=11 supergravity with those of D=10 IIA supergravity. Setting $y=X^\natural$,
the KK ansatz for the bosonic IIA string-frame fields is
\bea\label{twod}
ds^2_{11} &=& e^{-{2\over3}\phi(x)} ds^2_{10} + e^{{4\over3}\phi(x)} \big( dy+
dx^\mu C_\mu(x)\big)^2 \nn
A_{11} &=& A(x) + dy\wedge B(x) \, .
\eea
where we now suppose that $y$ is periodically identified with period $2\pi$.
The 10-metric, `dilaton' field $\phi$ and two-form potential
$B$ are the massless NS-NS fields of IIA superstring theory, while the D=10
gauge potentials $A$ and $C$ are the massless R-R fields. Defining
$G= dC$ one finds the following 6-brane solution of the IIA  supergravity field
equations \cite{HS}
\bea\label{twoe}
ds^2_{10} &=& H^{-{1\over2}}ds^2 (\bE^{(6,1)}) + H^{1\over2} d{\bf x}\cdot
d{\bf x}\nn 
e^{-4\phi} &=& H^3\nn
G &=& *dH\, ,
\eea
where $H$ is a harmonic function on $\bE^3$ (with coordinates ${\bf x}$) and $*$
indicates the Hodge dual on $\bE^3$. The singularities of $H$ are the positions
of parallel 6-branes; far away from these singularies the metric will be
asymptotic to the D=10 Minkowski metric if we choose $H$ such that $H\rightarrow
1$ as 
$|{\bf x}|\rightarrow\infty$.  From the ansatz (\ref{twod}) we can read off
the corresponding D=11 supergravity solution. It has $F=0$ and 11-metric 
\be\label{twof}
ds^2 = ds^2(\bE^{(6,1)}) + H\; d{\bf x}\cdot d{\bf x} + H^{-1}\; \big( dy + {\bf
C}\cdot d{\bf x}\big)^2\, . 
\ee
Given the assumed asymptotic property of $H$, the 11-metric is asymptotic to the
KK vacuum $\bE^{(9,1)}\times S^1$ as $|{\bf x}|\rightarrow\infty$, confirming 
the KK interpretation. 

If we set $|{\bf x}|=r$ and choose
\be\label{twog}
H= 1 + {1\over 2r}
\ee
then the 11-metric (\ref{twof}) is just the product of $\bE^{(6,1)}$ with the
Euclidean-Taub-NUT (ETN) 4-metric. The dimensional reduction to
D=10 yields the metric of a single 6-brane located at the origin
of $\bE^3$. However, whereas the 6-brane solution is genuinely singular at 
$r=0$, the corresponding singularity of the 11-metric is just a coordinate
singularity, provided that $y$ is identified with period $2\pi$. To see this we
note that $H\sim 1/2r$ near the singularity, so the metric near the
singularity is
\bea\label{twoh}
ds^2_4 &\sim & {1\over 2r} (dr^2 + r^2 d\Omega^2_2) + 2r(dy +C)^2\nn
&\sim & {dr^2\over 2r} + {r\over2}\big[ d\Omega^2_2 + 4(dy +C)^2\big]
\eea
where $d\Omega^2_2$ is the $SO(3)$-invariant metric on the 2-sphere.
Setting $\rho=\sqrt{2r}$ we have 
\be
ds^2_4 \sim  d\rho^2 + \rho^2 \big[ d\Omega^2_2 + 4(dy +C)^2\big]\; .
\ee
Now, provided that $y\sim y + 2\pi$, the hypersurfaces of constant $\rho$ are
3-spheres, each being a $U(1)$ bundle over $S^2$ (the Hopf fibration), so the
asymptotic metric near the singularity of $H$ is just the metric on $\bE^4$ in
polar coordinates. The `M-monopole' metric (\ref{twof}) therefore interpolates
between the `M-theory vacuum' $\bE^{(10,1)}$ near $r=0$ and the KK vacuum
$\bE^{(9,1)}\times S^1$ near $r=\infty$. The M-2-brane and M-5-brane solutions of
D=11 supergravity similarly interpolate between maximally-supersymmetric vacuum
solutions \cite{gibtown}, with the difference that the `M-theory vacuum' is at
now at infinity and one finds either the $adS_4\times S^7$ or the $adS_7\times
S^4$ KK vacuum near the singularities of $H$ (i.e. at the horizons).

Having now dealt with the significance of M-theory configurations carrying the
time-component of the 5-form charge $Y$ it remains for us to consider the
significance of the time component of the two-form charge $Z$. If we suppose
that $Z^{0\natural}=\mp P^0$ with all other charges vanishing then we find that
$\{Q,Q\} = P^0(1\mp \Gamma_\natural)$. Configurations preserving 1/2
supersymmetry are now associated with spinors $\epsilon$ satisfying the
constraint $\Gamma_\natural\epsilon =\pm \epsilon$. Again, this is not
immediately interpretable as the condition imposed by the presence of a brane
but it is equivalent to 
\be\label{symresb}
\Gamma^{0123456789}\epsilon =\pm\epsilon
\ee
which suggests a 9-brane. It seems that this must be interpreted as
a constraint associated with a (9+1)-dimensional boundary of the
11-dimensional spacetime, or `M-boundary', as in the Ho{\v r}ava-Witten (HW)
description of the heterotic string \cite{Horwit}. Note that the D=11 constraint
(\ref{symresb}) is equivalent to the D=10 chirality constraint
$\Gamma_{11}\epsilon=\pm\epsilon$, so the field theory on the
M-boundary boundary is indeed chiral. The D=10 interpretation of the M-boundary
is as the Minkowski vacuum of the heterotic string theory (which might equally
well be called the IIA 9-brane \cite{hullb}). The connection to {\it
bone fide} branes becomes clearer if we instead consider $Z^{09}=P^0$
but still compactify to the IIA theory in the $\natural\,$th direction. The
spinor constraint associated with 1/2 supersymmetric configurations is then 
equivalent to
\be\label{twoj}
\Gamma_{012345678}\Gamma_{11}\epsilon=\pm\epsilon\, ,
\ee
which is naturally associated with an 8-brane; in fact, it is the IIA
D-8-brane constraint \cite{Tasi}.

We have now discussed the significance of all charges in the M-theory algebra.
We have seen that the time components of the 2-form and 5-form charges can be
interpreted as those carried by branes of the $S^1$-compactified M-theory, alias
IIA superstring theory. All other IIA branes can be obtained directly from the
M-wave, M-2-brane and M-5-brane. Consider, for example, the M-wave. 
Compactifying M-theory in one of the nine transverse (${\bf x}$) directions we
obviously get a similar wave solution of IIA supergravity, but if we choose $K$
to be $u$-independent then we can also consider compactification in the direction
parametrized by $y= v-u$. Defining $t=v+u$, the M-wave 11-metric can then be
written as
\bea\label{twok}
ds^2_{11} &=& -dt^2 + dy^2 + K({\bf x})(dt-dy)^2 + d{\bf x}\cdot d{\bf x}\nn
&=& -H^{-{1\over2}}\big[H^{-{1\over2}} dt^2 - H^{1\over2} d{\bf x}\cdot d{\bf
x}\big] + H\big[ dy + (H^{-1}-1)dt\big]^2
\eea 
where
\be\label{twol}
H= 1+K\, .
\ee
We may now use the ansatz (\ref{twod}) to deduce that the following IIA
configuration is a solution of IIA supergravity
\bea\label{twom}
ds^2_{10} &=& -H^{-{1\over2}} dt^2 + H^{1\over2} d{\bf x}\cdot d{\bf x}\nn
e^{{4\over3}\phi}&=& H\nn
G &=& dt\wedge dH^{-1}\; .
\eea
Note that $H\rightarrow 1$ at infinity because $K\rightarrow 0$, so this
solution is asymptotically flat. In fact, it is a kind of extreme black hole
although the isolated singularities of $H$ are not horizons but rather genuine
singularities  of the metric, which must be resolved by going beyond IIA
supergravity. In the context of IIA superstring theory the solution 
(\ref{twom}) gives the long-range fields of the D-0-brane. We note here that 
the condition (\ref{onef}) imposed on Killing spinors by the presence of an
M-wave in D=11 becomes, in D=10, the condition
\be\label{twon}
\Gamma_0\Gamma_{11}\epsilon =\pm \epsilon\, ,
\ee
which is characteristic of a D-0-brane. Since a D-0-brane is 1/2 supersymmetric
its mass is simply related to its charge, which is quantized as a consequence of
the Dirac-Nepomechie-Teitelboim (DNT) quantization condition between 0-branes
and their 6-brane magnetic duals. In the context of M-theory this quantization
condition is due to the compactness of the 11th dimension. The quantization
condition on the 6-brane charge is then just what is needed to ensure the
non-singularity of the M-monopole metric in D=11, as discussed above.  

Let us now turn to the membrane solution (\ref{ones}). We can write it as
\bea\label{twona}
ds^2_{11} &=& H^{1\over3}\big[ H^{-1} ds^2(\bE^{(1,1)}) + ds^2(\bE^8)\big]
+ H^{-{2\over3}} dy^2 \nn
F &=& \big[vol(\bE^{(1,1)})\wedge dH^{-1}\big] \wedge dy
\eea
The ansatz (\ref{twod}) then leads to the following string solution of IIA
supergravity
\bea\label{twonb}
ds^2_{10} &=& H^{-1} ds^2(\bE^{(1,1)}) + ds^2(\bE^8)\nn
e^{-2\phi} &=& H\nn
F_{(3)} &=& vol(\bE^{(1,1)})\wedge dH^{-1}
\eea
where $F_{(3)}=dB$ is a 3-form field strength. This gives the long range fields
of  an infinite straight IIA superstring, which may therefore be
interpreted as the D=11 membrane wrapped on the KK circle \cite{DHIS}. 

If we choose the harmonic function in (\ref{ones}) to be independent of one of
the $\bE^8$ coordinates, which we now call $y$, then we can instead rewrite the
membrane 11-metric as
\be\label{twonc}
ds^2_{11} = H^{-{1\over6}}\big[ H^{-{1\over2}}ds^2(\bE^{(2,1)}) +
H^{1\over2}ds^2(\bE^7)\big] + H^{1\over3} dy^2
\ee
where $H$ is now harmonic on $\bE^7$. From the ansatz (\ref{twod}) we can then
read off a membrane solution of IIA supergravity. Note that it has a similar
form to the 6-brane and 0-brane solutions already given. In fact, all are
special cases of a general type II supergravity D-p-brane solution (even
$p$ for IIA and odd $p$ for IIB) for which the 10-metric and dilaton take the
form
\bea\label{twoo}
ds^2_{10} &=& H^{-{1\over2}} ds^2(\bE^{(p,1)}) + H^{1\over2} ds^2(\bE^{9-p})\nn
e^{4\phi} &=& H^{3-p}
\eea
where $H$ is a harmonic function on $\bE^{9-p}$. The metric is asymptotically
flat (for $p\le6$) if $H\rightarrow 1$ at infinity, and if we assume isolated
singularities then these can be interpreted as the positions of parallel
p-branes in static equilibrium. Clearly, the $p\ge7$ cases are special because
$H$ is logarithmic for $p=7$ and linear for $p=8$. The 8-brane case of
(\ref{twoo}) is a solution of the `massive' IIA supergravity theory which is
not directly obtainable by reduction of D=11 supergravity\footnote{This is
perhaps not unexpected in view of the association made above of the 8-brane
charge with a D=11 spacetime boundary because a boundary is determined by
imposing boundary conditions rather than by solving local field equations.
Perhaps the distinction between these two aspects of traditional physics is
blurred in M-theory.}. 

Turning finally to the 5-brane solution (\ref{onev}) we may either wrap it on
$S^1$, which yields the D-4-brane solution, i.e. the $p=4$ case of (\ref{twoo}),
or we may simply reduce it to D=10, as done above for the membrane, to get the
following IIA 5-brane solution
\bea\label{twop}
ds^2_{10} &=& ds^2(\bE^{(5,1)}) + H ds^2(\bE^4) \nn
e^{2\phi} &=& H \nn
F_{(3)}&=& * dH
\eea
where $H$ is now harmonic on $\bE^4$ and $*$ indicates the Hodge dual on
$\bE^4$. This is the magnetic dual of the IIA string solution. Unlike the string
solution, however, the dual 5-brane solution is non-singular in the sense that
the metric is geodesically complete, assuming of course that $H$ has only point
singularities and is such that
$H\rightarrow 1$  at infinity. Near a singularity at $r=0$ in $\bE^5$, where $r$
is a radial coordinate, we have $H\sim 1/r^3$ so that the distance to the
singularity from
$r=R$ is
\be\label{twoq}
ds \sim  \int_0^R {dr\over r^{3/2}} = \infty\, .
\ee
This is the distance on a spacelike geodesic. Because of the direct product
structure of the metric it suffices to consider spacelike geodesics. Thus, the
singularities of $H$ are at infinite affine parameter on all geodesics and
the IIA 5-brane {\sl string-frame} metric is therefore geodesically complete
\cite{HS}. As the 5-brane is the magnetic dual of the `fundamental' string it is
perhaps not surprising that it should be non-singular. Note, however, that the
non-singularity is achieved by a very different mechanism to that of the
M-5-brane, for which the singularities of $H$ are horizons at finite affine
parameter. 

We have now seen all the p-brane solutions of D=11 supergravity and IIA D=10
supergravity. The relation between them is nicely summarized by the reduction
to D=10 of the M-theory superalgebra. Simply decomposing the $SO(10,1)$
representations into $SO(9,1)$ representations we obtain the algebra
\bea\label{twor}
\{Q_\a,Q_\b\} &=& (C\Gamma^\mu)_{\a\b} P_\mu + (C\Gamma_{11})_{\a\b} P_{11} 
+ (C\Gamma^\mu\Gamma_{11})_{\a\b} Z_\mu \nn
&& + {1\over2}(C\Gamma^{\mu\nu})_{\a\b} Z_{\mu\nu} +
{1\over4!}(C\Gamma^{\mu\nu\rho\sigma}\Gamma_{11})_{\a\b}Y_{\mu\nu\rho\sigma} \nn
&& +\, {1\over 5!} (C\Gamma^{\mu\nu\rho\sigma\lambda})_{\a\b}
Y_{\mu\nu\rho\sigma\lambda}\, .
\eea 
We can associate the space components of $P_\mu$ with IIA-waves, or
supergravitons of IIA supergravity. The central charge $P_{11}$ is carried by
D-0-branes which (together with their threshold bound states) are 
IIA-superstring manifestation of the KK-modes arising from the compactification
from D=11 \cite{revisited,Witvarious}. Each space component of $Z_\mu$ is a
charge carried by a `fundamental' IIA superstring, the component depending upon
the orientation of the string. All other charges are similarly carried by other
IIA p-branes.

We conclude this lecture with a brief digression from the main line of our
argument. Recall that {\sl isolated} singularities of the harmonic function $H$
in the M-KK-monopole 11-metric (\ref{twof}) are just coordinate singularities,
although they are genuine singularities of the IIA 6-brane 10-metric. The
singularity in D=10 can  be viewed as arising from an illegitimate neglect of
the 11th dimension. But what about {\sl non-isolated} singularities? Consider 
the harmonic function
\be\label{symresa}
H= 1 + { 1\over 2|{\bf x} -{\bf a}|} + {1\over 2|{\bf x} +{\bf a}|}\, ,
\ee
which represents two `parallel' M-KK-monopoles, i.e. that D=11 supergravity
configuration reducing in D=10 to two parallel IIA 6-branes. At the two
singularities of $H$ the KK circle contracts to a point. The two M-KK-monopoles
are therefore connected by a homology 2-sphere with azimuthal angle
$y$ and poles at ${\bf x}=\pm {\bf a}$. If we now take $|{\bf a}|\rightarrow 0$
then this homology 2-sphere shrinks to a point and the metric acquires a
genuine curvature singularity. 

To resolve this singularity one needs to take 
into account the fact that membranes may wrap around the 2-sphere. The total
energy of such a membrane is proportional to the area $A$ of the 2-sphere, at
least for the large area limit in which the semiclassical description of the
M-2-brane is valid. One would normally expect this semiclassical result to be
modified by corrections that are small for large $A$ but become dominant as
$A\rightarrow 0$. However, supersymmetry ensures that this does not happen, and
wrapped M-2-branes become massless as $A\rightarrow 0$ \cite{hulltown}. This
was originally shown for $K_3$ compactifications of M-theory,
following the suggestion in \cite{Witvarious} that the symmetry enhancement
expected on the basis of duality with the heterotic string should be associated
with collapsing 2-cycles of $K_3$. But the geometry near a collapsing two-cycle
of $K_3$ is the same as the geometry near a pair of nearly-coincident parallel
M-theory monopoles\footnote{This point has been made independently by Sen
\cite{Sen}.}. The geometry in the $K_3$ case is that of the Eguchi-Hanson
instanton which differs from two-centre ETN metric only by the absence of the
`$1$' term in the harmonic function $H$. This difference is insignificant near
the centres of the metric when $|{\bf a}|\rightarrow 0$. From the D=10
perspective, the wrapped membranes are strings stretched between two D-6-branes
and the massless states that appear in the coincidence limit are the string
states that lead to an enhancement to $U(2)$ of the $U(1)\times U(1)$ gauge
symmetry on the 6-branes's common worldvolume \cite{Wittenb}. Thus, the short
distance, or UV, singularity of the classical supergravity solution that occurs
when $|{\bf a}|=0$ is due to an illegitimate neglect of massless supermembrane
states. The inclusion of these additional massless  states resolves the
singularity. We see that at least some of the UV singularities of supergravity
are resolved in M-theory by relating them to the IR physics of massless
particles coming from membranes. The resolution of UV singularities of quantum
gravity by new IR physics on branes is embodied in the M(atrix) model approach
to M-theory, which will not be discussed here (and for which we refer to
\cite{Banks} for a recent review). Instead we shall continue to concentrate on
what can be learnt from classical solutions of D=11 supergravity and, more
abstractly, the M-theory superalgebra.

\section{Lecture 3: Dualities}

There is a web of dualities connecting M-theory with both the Type IIA and Type
IIB superstring theories, a summary of which may be found in my previous
lectures on M-theory \cite{fourM}. The relation between M-theory and the Type
IIA superstring theory will be called `M-duality'; we have just seen how some
aspects of this relation are encoded in the respective supersymmetry algebras. 
We are now going to see how the `T-duality' connecting the IIA and IIB
superstring theories is similarly encoded. 

We begin by reconsidering the IIA algebra of (\ref{twor}) in a form in which 
the D=10 Majorana supercharge $Q$ is decomposed into the sum of two
Majorana-Weyl supercharges $Q^\pm$ of opposite chirality, i.e. 
\be\label{threea}
Q^\pm = {\cal P}^\pm Q \, \qquad {\cal P}^\pm \equiv 
{1\over2}(1\pm \Gamma_{11})\, .
\ee 
The IIA supersymmetry algebra (\ref{twor}) becomes
\bea\label{threeb}
\{ Q^+_\a,Q^+_\b\} &=& (C{\cal P}^+ \Gamma^\mu)_{\a\b} (P + Z)_\mu +
{1\over 5!} (C\Gamma^{\mu\nu\rho\sigma\lambda})_{\a\b}
Y^+_{\mu\nu\rho\sigma\lambda} \nn
\{ Q^-_\a,Q^-_\b\} &=& (C{\cal P}^- \Gamma^\mu)_{\a\b} (P - Z)_\mu +
{1\over 5!} (C\Gamma^{\mu\nu\rho\sigma\lambda})_{\a\b}
Y^-_{\mu\nu\rho\sigma\lambda} \nn
\{Q^+_\a ,Q^-_\b\} &=& (C{\cal P}^+ )P_{11} +
{1\over2}(C{\cal P}^+\Gamma^{\mu\nu})_{\a\b} Z_{\mu\nu} \nn
&& +\, 
{1\over 4!}(C{\cal P}^+\Gamma^{\mu\nu\rho\sigma})_{\a\b}Y_{\mu\nu\rho\sigma}\, ,
\eea
where $Y^\pm$ are (anti)self-dual 5-form charges. The p-forms occuring in the
last anticommutator are the charges carried by the D-p-branes of type IIA
superstring theory. There are manifestly charges for $p=0,2,4$, but there are
also charges for $p=6,8$ coming from the time components of
$Y_{\mu\nu\rho\sigma}$ and $Z_{\mu\nu}$, respectively. These D-brane charges
couple to the massless R-R fields of IIA superstring
theory and so are also called the (IIA) `R-R charges'. 
All other charges are `NS-NS charges' because they couple to massless
fields in the  NS-NS sector of the superstring theory.
The latter include the 1-form charge carried by the IIA superstring itself and
{\sl two} 5-form charges, counting the algebraically irreducible self-dual
and anti-self-dual 5-forms separately. The combination $Y=Y^+ + Y^-$ is the
5-form charge descending from the 5-form in D=11 and is therefore the charge
carried by the magnetic 5-brane dual to the IIA string. This is called the
NS-5-brane or the `solitonic' 5-brane (S-5-brane). The other combination $\tilde
Y= Y^+ -Y^-$ is the 5-form charge associated to IIA-KK-monopoles. 

The IIA algebra is invariant under the transformation for which $Q^-$ and all RR
charges change sign. If we `mod out' by this symmetry we arrive at the N=1 D=10
supersymmetry algebra of the heterotic string (since this is the N=1
superstring without a RR sector). This heterotic supersymmetry algebra, 
is equivalent to
\be\label{threec}
\{ Q^+_\a,Q^+_\b\} = (C{\cal P}^+ \Gamma^\mu)_{\a\b} (P + Z)_\mu 
 +  {1\over 2.5!} (C\Gamma^{\mu\nu\rho\sigma\lambda})_{\a\b} (Y +
\tilde Y)_{\mu\nu\rho\sigma\lambda}\, .
\ee
Note that this is invariant under the interchange of, say, $P_9$ with
$Z_9$. For this to represent a symmetry of the heterotic string theory the
spectra of these two operators would have to coincide. This is not normally the
case but if the $9$-direction is a circle, of radius $R$, then the spectra of
$P_9$ and $Z_9$ are isomorphic, the isomorphism involving the transformation
$R\rightarrow 1/R$. In other words, the spectrum of the heterotic string theory
compactified on a circle of radius
$R$ is identical to that of the same string theory compactified on a circle of
radius $1/R$ because the transformation $R\rightarrow 1/R$ exchanges the KK
modes, i.e. the spectrum of $P_9$, with the string winding modes, i.e. the
spectrum of $Z_9$. In fact, the two heterotic string theory compactifications
are equivalent to all orders in string perturbation theory; they are said to be
`T-dual'. 

To see whether this remains true of the fully non-perturbative theory one must
consider the effects of the T-duality transformation on the non-perturbative
spectrum. For example, having chosen $X^9$ as the parameter of a KK circle of
radius $R$, we have KK-monoples with p-form charges $\tilde Y$ proportional to
$R$. We also have a KK tower of 5-branes with charges $Y$ proportional to $1/R$.
Now, just as compactification from D=11 to D=10 allows the M-KK-monopole to be
interpreted as a 6-brane, so compactification from D=10 to D=9 allows a D=10
KK-monopole to be interpreted as a 5-brane. This is nominally a 5-brane in D=9
but it can be re-interpreted in D=10 as one of the KK tower of 5-branes. Given
that the KK-monopole charge was, say, $\tilde Y_{06789}$, the charge on the
T-dual 5-brane will be $Y_{12345}$. In fact, for each KK-monopole with charge
$\tilde Y = {1\over2}(Y^+ -Y^-)$ we have a T-dual 5-brane with charge $Y =
{1\over2}(Y^+ + Y^-)$, so at least this sector of the non-perturbative spectrum
is invariant under $R\rightarrow 1/R$ provided that we also 
take $Y^- \rightarrow - Y^-$ (for all $09$ and 8-space components). Since only
$Y^+$ appears in the heterotic algebra (\ref{threec}), this algebra is 
invariant, consistent with the non-perturbative validity of heterotic T-duality. 

So far, T-duality can be summarized by the statement that
T-duality in the 9-direction effects the transformation
\be\label{threed}
(P-Z)_9 \rightarrow -(P-Z)_9, \qquad Y^-_{0abcd9}\rightarrow - 
Y^-_{0abc 9},\qquad Y^-_{abcde} \rightarrow - Y^-_{abcde}
\ee
where $a,b,c,d,e = 1,2,\dots, 8$. The IIA superstring theory cannot be T-dual to
itself because the full IIA supersymmetry algebra is definitely not invariant
under this transformation. Of course, in performing this exchange we are just
relabelling the generators that span the algebra, but it is a relabelling that
destroys the manifest D=10 Lorentz covariance. One might suppose that manifest
Lorentz covariance could be restored only by reversing the interchange but,
remarkably, there exists another way. We first note that (\ref{threed}) changes
some signs, in a non-Lorentz covariant fashion, in the $\{Q^-,Q^-\}$
anticommutator. We can reverse these sign changes by defining a new charge
$\tilde Q^+$ by
\be\label{threee}
Q^- =  \tilde Q^+ \Gamma_9\, .
\ee
The new charge is chiral rather than antichiral because multiplication by
$\Gamma^9$ changes the chirality. We now find that 
\be\label{threef}
\{ \tilde Q^+ _\a,\tilde Q^+ _\b\} = (C{\cal P}^+ \Gamma^\mu)_{\a\b} (P -
Z)_\mu + {1\over 5!} (C\Gamma^{\mu\nu\rho\sigma\lambda})_{\a\b}
V^+_{\mu\nu\rho\sigma\lambda}
\ee
where $V^+$ is a new self-dual 5-form such that
\be\label{threeg}
V^+_{mnpqr} = Y^-_{mnpqr} \qquad (m,n,p,q,r = 0,1,\dots,8),
\ee
the other components being determined by (anti)self-duality.
The remaining anticommutator with the RR charges becomes 
\bea\label{threeh}
\{\tilde Q^+_\a,\tilde Q^+_\b\} &=& (C{\cal P}^+\Gamma^\mu)_{\a\b} \tilde Z_\mu
+  {1\over 3!} (C{\cal P}^+\Gamma^{mu\nu\rho})_{\a\b}W_{\mu\nu\rho}\nn
&& + {1\over 5!} (C\Gamma^{\mu\nu\rho\sigma})_{\a\b}
\tilde V^+ _{\mu\nu\rho\sigma\lambda} 
\eea
where 
\bea\label{threei}
\tilde Z_\mu &=& (Z_{m9},P_{11})\nn
W_{\mu\nu\rho} &=& (Y_{mnp9},Z_{mn})\nn
\tilde V^+_{mnpq9} &=& Y_{mnpq}\, .
\eea

We have now arrived at the other D=10 N=2 supersymmetry algebra, the IIB
algebra, which has two chiral supercharges $Q^I =(Q^+,\tilde Q^+)$. 
The $\{Q,Q\}$ anticommutator can be written in the form
\bea\label{threek}
\{ Q^I_\a,Q^J_\b\} &=& (C{\cal P}^+\Gamma^\mu)_{\a\b}\big( \delta^{IJ} P_\mu 
+ \sigma_3^{IJ} Z_\mu + \sigma_1^{IJ}  \tilde Z_\mu\big) \nn
&& + {1\over 3!} \sigma_2^{IJ}\big( C{\cal P}^+\Gamma^{\mu\nu\rho})_{\a\b}
W_{\mu\nu\rho} + {1\over 5!} \delta^{IJ}
(C\Gamma^{\mu\nu\rho\sigma\lambda})_{\a\b}K^+_{\mu\nu\rho\sigma\lambda} \nn &&+
{1\over 5!} (C\Gamma^{\mu\nu\rho\sigma\lambda})_{\a\b}
\big(\sigma_3^{IJ} V^+_{\mu\nu\rho\sigma\lambda} +
\sigma_1^{IJ}\tilde V^+_{\mu\nu\rho\sigma\lambda}\big)
\eea
where
\be\label{threel}
K^+ = Y^+ - V^+ \, .
\ee
The process by which we arrived at this algebra suggests, correctly
\cite{dine,leigh}, that the T-dual of the IIA superstring theory is the IIB
superstring theory, and vice-versa. T-duality relates the IIA theory
compactified on a circle of radius $R$ to the IIB-theory compactified on a circle
of radius $1/R$. Of course, neither theory is D=10 Lorentz covariant for finite
$R$ or $1/R$, but D=10 Lorentz covariance is recovered in either of the two
limits $R\rightarrow 0$ or $R\rightarrow \infty$. {\sl We also learn from the
T-duality map between the two Type II algebras how the branes of one theory are
to be interpreted in the other one}. 

For example, from the fact that the transverse 1-form charge $Z_m$ in the IIA
algebra is mapped to the same charge in the IIB algebra we learn that T-duality
in a direction perpendicular to the IIA string transforms it into the IIB
string, and vice-versa. This can be verified by inspection of the solutions
representing the long-range fields of the IIA and IIB strings. In fact, the
solution is the same for both IIA and IIB, and it is also a solution of the
effective supergravity field equations of the heterotic string. The
(string-frame) 10-metric of this common string solution is given in
(\ref{twonb}). T-duality in a direction perpendicular to  the string requires
that we compactify one of the directions of the transverse 8-space, which
becomes $\bE^7\times S^1$. The absence of any power of the harmonic function
multiplying the metric on this transverse space shows that the 10-metric is
invariant under the inversion of the radius of the
$S^1$ factor, apart from a possible constant rescaling of the coordinates to
achieve a standard identification of the $S^1$ coordinate. In the heterotic
case we interpret this to mean that the heterotic string is mapped into itself.
In the Type II cases we must interpret it to mean that the IIA string is mapped
to the IIB string, and vice-versa. 
Similarly, the fact that $Z_9$ in the IIA(B) algebra is mapped to $P_9$ in
the IIB(A) algebra shows that T-duality in a direction parallel to a IIA(B)
string results in a IIB(A)-wave. This is slightly less straightforward to verify
at the level of supergravity solutions, essentially because the wave has an
off-diagonal metric. Since all fields involved are common to the Type II and
heterotic strings the details can be found in most other accounts of T-duality 
and will therefore be omitted here.

Turning to the 5-form charges in the Type II algebras we see that T-duality of
a IIA-KK-monopole in a `parallel' direction yields a IIB-KK-monopole, where 
`parallel' means in a direction other than that `occupied' by the
ETN 4-metric of the KK-monopole. This can also be seen from
inspection of the IIA-KK-monopole metric. The D=11 KK-monopole metric is given
in (\ref{twof}). Using the dimensional reduction ansatz (\ref{twod}) we see that
the IIA dilaton is constant and the IIA 10-metric is 
\be\label{threej}
 ds^2_{10} = ds^2(\bE^{5,1}) + H d{\bf x}\cdot d{\bf x} + H^{-1}(dy+ C)^2\; . 
\ee
The only non-zero fields in the complete solution are those in common with 
IIB supergravity, so the same solution also serves as the IIB-KK-monopole.
The absence of any power of the harmonic function multiplying the 6-dimensional
`worldvolume' factor confirms that T-duality in these directions just takes the
IIA-KK-monopole into the IIB-KK-monopole. If, on the other hand, we T-dualize
in the $y$-direction then the $H^{-1}$ factor is inverted and we begin to see the
emergence of a 5-brane metric. This is confirmed by inspection of the T-duality
map between the 5-form charges in the Type II algebras. 

We now have a more or less complete set of rules for T-duality of objects
carrying NS-NS charges. The results are essentially the same as those for the
heterotic string with the difference that each time we T-dualize we move from
the IIA(B) to the IIB(A) theory. The principal novelty of T-duality in the Type
II context is its effect on the D-branes carrying the RR charges \cite{Tasi}.
When the T-duality map (\ref{threei}) between RR charges is interpreted in terms
of branes we see, for example, that T-duality in a direction parallel to a
D-p-brane results in a $D-(p-1)$-brane in the dual theory. This can be understood
as follows. In performing the T-duality we first compactify on a circle of large
radius $R$. Saying that this direction is parallel to the p-brane amounts to
saying that the p-brane is wrapped around this direction. As we take
$R\rightarrow 0$ the p-brane becomes, effectively, a $(p-1)$-brane. {\sl A
priori}, we might expect to have to interpret this as a $(p-1)$-brane in a D=9
theory but T-duality allows us to re-interpret it as a $(p-1)$-brane in the
T-dual D=10 theory. Conversely, T-duality in a direction orthogonal to a RR
$p$-brane results in a RR $(p+1)$-brane in the T-dual theory. We could also have
considered T-duality in a direction that is neither parallel nor perpendicular
to a given brane, but we shall ignore this (inessential) complication here. 

The T-duality map between RR p-branes in the IIA and IIB theories is nicely
reflected in the form of the 10-metric (\ref{twoo}) of the corresponding
supergravity solutions. A T-duality transformation takes the radius of a circle
to the inverse radius. If one supposes that the circle in question is
parameterized by one of the cartesian space coordinates of either
$\bE^{(p,1)}$ (i.e. a direction parallel to the p-brane) or $\bE^{9-p}$ (i.e. a
direction orthogonal to the p-brane) then the inversion of the radius either
takes one factor of $H^{-{1\over2}}$ to $H^{1\over2}$ or vice-versa,
respectively. Thus the $p$-brane metric is taken either to the $(p-1)$-brane
metric or to the $(p+1)$-brane metric. We have still to consider the other fields
of the D-brane supergravity solutions but these merely confirm the result
suggested by the T-duality transformation of the metric, which is the same
result as we deduced above from the T-duality map between the Type II
supersymmetry algebras. 

We are now in a position to see how some of the various branes of M-theory and
Type II string theories are related to each other by dualities. Let us start
from an M-wave in the $\natural$ direction. It is convenient to represent this by
the array
$$
\begin{array}{lcccccccccc}
MW:& - &- & - & - & - & - & - & - & - & \natural 
\end{array}
$$
Reduction to D=10 in the $\natural$ direction yields the D-0-brane, which we
represent by the array
$$
\begin{array}{lcccccccccc}
D0: & - & - & - & - & - & - & - & - & - 
\end{array}
$$
T-duality in the 1-direction now yields the D-string oriented in the 1-direction
$$
\begin{array}{lcccccccccc}
D1: & 1 & - & - & - & - & - & - & - & - 
\end{array}
$$
Further T-duality in the 2-direction yields the D-2-brane
$$
\begin{array}{lcccccccccc}
D2: & 1 & 2 & - & - & - & - & - & - & - 
\end{array}
$$
This can be lifted to D=11 to the M-2-brane
$$
\begin{array}{lccccccccccc}
M2: & 1 & 2 & - & - & - & - & - & - & - & -
\end{array}
$$
Instead, we may continue to T-dualize the D-2-brane until arriving at the
D-4-brane
$$
\begin{array}{lccccccccc}
D4: & 1 & 2 & 3 & 4 & - & - & - & - & - 
\end{array}
$$
which we may then lift in the $\natural$ th direction to arrive at an 
M-5-brane in the 1234$\natural$ 5-plane
$$
\begin{array}{lccccccccccc}
M5: & 1 & 2 & 3 & 4 & - & - & - & - & - & \natural
\end{array}
$$
If we instead continue to T-dualize to the D-6-brane
$$
\begin{array}{lcccccccccc}
D6: & 1 & 2 & 3 & 4 & 5 & 6 & - & - & - 
\end{array}
$$
then we may lift in the $\natural$ th direction to an M-KK-monopole. We
represent this by the array
$$
\begin{array}{lcccccccccc}
MKK: & 1 & 2 & 3 & 4 & 5 & 6 & - & - & - & \times 
\end{array}
$$
where the cross indicates the compact KK-circle of the KK-monopole. 
In carrying out these steps at the level of supergravity solutions we
start from the harmonic function $K$ of the M-wave. At the first step this
is converted into the harmonic function $H$ of the D-0-brane in the way
indicated earlier, and this harmonic function then appears in all the
subsequent dualized solutions. However, certain steps require $H$ to be
independent of the coordinate in the T-duality direction. For example, in 
passing from the D-0-brane to the D-2-brane $H$ goes from a harmonic function
on $\bE^9$ to one on $\bE^7$.  Re-interpreting this as a solution of D=11
supergravity then yields the M-2-brane solution but with $H$ harmonic on
$\bE^7$ instead of $\bE^8$. On the other hand, if we instead continue to
T-dualize to the D-4-brane then $H$ is reduced to being harmonic on $\bE^5$
but when the D-4-brane solution is re-interpreted in D=11 we recover the
general M-5-brane solution. 

The D-6-brane is the magnetic dual of the D-0-brane. If we continue to
T-dualize we arrive at the IIB 7-brane and IIA 8-brane (the `high-branes')
for which there are no obvious electric duals. The asymptotic behaviour of the
7-brane and 8-brane supergravity solutions are also special. In the 7-brane
case the function $H$ is harmonic on $\bE^2$, so point singularities are conical
singularities and the energy density per unit 7-volume is logarithmically
divergent. This case will not be discussed in these lectures. In the 8-brane 
case $H$ is harmonic on $\bE^1$; allowing for point singularities means that it
is piecewise linear. The 8-brane configuration of IIA supergravity is actually
not a solution of the standard IIA field equations but rather of a `massive'
variant with a cosmological constant. The 8-brane is effectively a IIA domain
wall separating regions of different cosmological constant \cite{polwit,green}.
Formally, we may continue to T-dualize to arrive at the IIB D-9-brane. The
function $H$ is now constant, so the D-9-brane solution is just flat D=10
Minkowski space. It might appear from this that, unlike all other D-branes, the
D-9-brane does not break 1/2 supersymmetry. But we still have the constraint
\be\label{athreek}
\Gamma^{0123456789}\epsilon =\pm \epsilon
\ee 
which is equivalent to the chirality constraint $\Gamma_{11}\epsilon =\pm
\epsilon$. It follows that the D-9-brane must be
interpreted as the Minkowski vacuum spacetime of a D=10 string theory with N=1
supersymmetry. Since the D-9-brane is a D-brane, strings may end on it. These
strings can only be those of the Type I string theory, because this is the only
D=10 superstring theory with both closed and open strings. 

One potential problem with the attempt to interpret the Type I string theory in
this way  is that the 8-brane carries a RR charge, for which the lines of force
must either terminate on an anti 8-brane or go off to infinity. As long as the
one transverse direction is non-compact the force lines may go off to infinity
but to T-dualize in this last direction we must first compactify it. Since the
lines of force cannot now wander off to infinity the total 8-brane charge must
vanish. If D-branes were the only objects to carry RR charge then the net number
of 8-branes minus anti 8-branes would have to vanish. Such a configuration would
relax to one with neither 8-branes nor anti 8-branes because the anti 8-branes
would attract the 8-branes and annihilate them. In this case the initial IIA
configuration preserving supersymmetry would be a IIA KK-vacuum, for which the
T-dual is a IIB KK vacuum, rather than a D-9-brane. Thus, to arrive at the Type I
string theory by T-duality of the IIA theory 8-brane requires an additional
ingredient. This ingredient, the orientifold 8-plane, is best understood from 
the other direction, i.e. by T-duality of the Type I string \cite{leigh}. 

The existence of the Type 1 string is suggested by the IIB supersymmetry algebra
since there are {\sl two} ways to truncate the latter to an N=1 supersymmetry
algebra, corresponding to the following two involutions of the IIB
algebra:
\bea\label{invola}
Q^+ &\rightarrow& \sigma_3 Q^+ \nn
\tilde Z &\rightarrow& -\tilde Z \nn
W &\rightarrow& -W \nn
\tilde V^+ &\rightarrow& -\tilde V^+
\eea
and
\bea\label{involb}
Q^+ &\rightarrow& \sigma_1 Q^+ \nn
\tilde Z &\rightarrow& -Z \nn
W &\rightarrow& -W \nn
\tilde V^+ &\rightarrow& -V^+\, .
\eea
We can obtain an N=1 supersymmetry algebra by `modding out' by either of these
involutions. In the first case we set $\tilde Q^+=0$ to recover the heterotic
string algebra of (\ref{threec}). In the second case, we must set $Q^+=\tilde
Q^+$. We then find an algebra isomorphic to (\ref{threec}) but with the NS
1-form $Z$ replaced by the RR 1-form $\tilde Z$ and the NS self-dual 5-form 
$Y^+$ replaced by the RR self-dual 5-form $\tilde V^+$. The fact that there is
now no NS string charge means that what was originally the fundamental type IIB
string has become a string without the 2-form charge needed to prevent it from
breaking. This fits the description of the Type 1 string. To confirm it we
must examine the consequences of the $Z_2$ identification on the IIB string
worldsheet fields.

If the IIB supercharges are expressed as Noether charges in terms of worldsheet
fields of the IIB superstring, then the $Z_2$ transformation (\ref{involb}) is
effected by a worldsheet parity transformation $\sigma\rightarrow -\sigma$. This
is a symmetry of the IIB superstring action provided all worldsheet scalars
$X^\mu$ are indeed scalars, rather than pseudo-scalars. Thus, the identification
under worldsheet parity leads to the Type 1 supersymmetry algebra and the
worldsheet scalar fields are now subject to $X^\mu(\sigma)=X^\mu(-\sigma)$.
If we now T-dualize in, say, the 9-direction then $X^9$ becomes a pseudoscalar
and the identification under worldsheet parity leads to $X^9(-\sigma) =
-X^9(\sigma)$. In particular $X^9(0)=-X^9(0)$ so the KK circle in the
9-direction is actually the interval $S^1/\bZ_2$. The 8-planes $X^9(0)=0$ are
effectively boundaries of the 9-dimensional space of the D=10 string
theory T-dual to the Type 1 string theory\footnote{This has been called the Type
I' theory but a possibly better terminology is Type IA, in which case the Type I
string could be renamed Type 1B.}. These are the orientifold 8-planes. Open
strings, which previously had endpoints in spacetime with $SO(32)$ Chan-Paton
factors, now end on any of 16 8-branes. The RR charge of these 8-branes is
cancelled by the RR charge of the orientifold planes. Much more could be said
about the Type 1 string theory but we have now learnt what we can from the
supersymmetry algebra.

We have one more duality transformation to consider before we can relate all
M-theory or Type II branes to each other. This is IIB S-duality
\cite{unity,Witvarious}. We first note that the IIB algebra is invariant under
the cyclic group generated by the transformation
\be\label{threem}
Q \rightarrow e^{{i\pi\over 4}\sigma_2} Q
\ee
on the two supercharges, combined with the transformation 
\be\label{threen}
(Z,X) \rightarrow (X,-Z) \qquad (V,\tilde V)\rightarrow (\tilde V,-V)
\ee
on the bosonic charges. This group is a subgroup of an $Sl(2,R)$
symmetry group of IIB supergravity. We see from its action on the IIB charges
that it interchanges the IIB string with the D-string and the IIB solitonic
5-brane with the D-5-brane. The IIB 3-brane is invariant. The group is
actually of order 4 because it squares to $-1\in Sl(2,R)$ but this is equivalent
to the identity element of $PSl(2,R)$, which is the isometry group of the target
space of the scalar fields of IIB supergravity. Thus, there is a $Z_2$ action on
the moduli space of the IIB theory exchanging NS-branes with RR-branes. This is
clearly non-perturbative within the context of Type IIB superstring theory, but
is believed to be a $Z_2$ subgroup of a $PSl(2,Z)$ S-duality group of the fully
non-perturbative IIB theory\footnote{It is possible that there exists a new
superstring theory containing both the IIB string and the D-string in which the
$PSl(2,Z)$ symmetry is visible in perturbation theory \cite{ceder}.}.
Here we shall need only the $Z_2$ subgroup which we shall also call S-duality. 

The effects of S-duality on IIB supergravity solutions are most easily seen in
the Einstein-frame because the Einstein-frame metric (defined by the absence of
a power of the dilaton multiplying the $\sqrt{-g}R$ term in the action) is
$Sl(2;R)$ invariant. The relation between the Einstein-frame metric $ds^2_E$ and
the string frame metric is
\be\label{threeo} 
ds^2 = e^{{1\over2}\phi} ds^2_E\, .
\ee
An S-duality transformation is then achieved by changing the sign of the 
dilaton while simultaneously exchanging the two 3-form field strengths in the 
way indicated by the exchange (\ref{threen}) of the 1-form charges. Thus, the
effect of S-duality on the string-frame metric and dilaton is
\be\label{threep}
ds^2 \rightarrow e^{-\phi}ds^2\; ,\qquad \phi\rightarrow -\phi\, .
\ee
Let us consider the effect of an S-duality transformation on the IIB string
solution, which is the same as the IIA string solution of (\ref{twonb}). In
particular, the metric and dilaton are given by 
\bea\label{threeq}
ds^2_{10} &=& H^{-1} ds^2(\bE^{(1,1)}) + ds^2(\bE^8)\nn
e^{-\phi} &=& H^{1\over2}\, .
\eea
The transformation (\ref{threep}) then yields
\bea\label{threer}
ds^2_{10} &=& H^{-{1\over2}} ds^2(\bE^{(1,1)}) + H^{1\over2} ds^2(\bE^8)\nn
e^{-\phi} &=&  H^{-{1\over2}}
\eea
which is precisely the metric and dilaton of the D-string. We thus confirm the
previous conclusion that IIB S-duality takes the IIB string to the D-string.
One can similarly confirm that IIB S-duality takes the IIB NS-5-brane to the
D-5-brane. 

By means of the three dualities, `M', `T' and `S', we can go, in the way
described above, from any of the 1/2-supersymmetric `objects' of M-theory or
Type II superstring theories to any other one. M-theory is therefore a theory
of a single object with various dual manifestations, no one of which
is sufficient by itself. The Type 1 and heterotic string theories might
at first appear to stand apart, but the objects within these theories preserving
1/2 of the supersymmetry of the D=10 N=1 Minkowski vacuum can be viewed as
special cases of 1/4 supersymmetric `intersecting brane' configurations of
M-theory or Type II superstring theory. For example, the D-5-brane of Type I
superstring theory can be viewed as a IIB D-5-brane inside a IIB 9-brane. As
another example, the heterotic string can be viewed as the boundary of an
M-2-brane on an M-boundary. Neither of these cases is, strictly speaking,
an example of `intersecting' branes but it is convenient to consider them under
this rubric because they are dual (at least formally) to cases to which the term
has its obvious meaning. We shall now consider some of these cases in more
detail.

\section{Lecture 4: Intersecting M-branes}

Each of the `basic', 1/2 supersymmetric, objects of M-theory or Type II
superstring theory, with a given orientation, is associated with a constraint of
the form $\Gamma\epsilon=\epsilon$ for some traceless product $\Gamma$ of Dirac
matrices with the property that $\Gamma^2=1$. Given two such objects we have
two matrices with these properties. Let us call them $\Gamma$ and
$\Gamma'$. Let $\zeta$ and $\zeta'$ be the charge/tension ratios of the objects
associated, respectively, with $\Gamma$ and $\Gamma'$. Then the $\{Q,Q\}$
anticommutator takes the form
\be\label{foura}
\{Q,Q\} = P^0\big[ 1 + \zeta \Gamma + \zeta' \Gamma'\big] \, .
\ee
Positivity imposes a bound on $\zeta$ and $\zeta'$, but the form of this bound
depends on whether the two matrices $\Gamma$ and $\Gamma'$ commute or
anticommute. 

If $\{\Gamma,\Gamma'\}=0$ then 
\be\label{fourb}
(\zeta \Gamma + \zeta' \Gamma')^2 = \zeta^2 + \zeta'{}^2\, ,
\ee
so the bound is $\zeta^2 + \zeta'{}^2 \le 1$, which is equivalent to a bound of
the type
\be\label{fourc}
T \ge \sqrt{Z^2 + Z'{}^2} 
\ee
where $Z$ and $Z'$ are the charges of the two branes. Since the right hand side
is strictly greater than $|Z|+|Z'|$, unless either $Z$ or $Z'$ vanishes, a
configuration saturating this bound must be a `bound state' with strictly
positive binding energy. It is associated with a constraint of
the form $\Gamma''\epsilon=\epsilon$ where
\be\label{fourd}
\Gamma '' = \cos\vartheta\, \Gamma + \sin\vartheta\, \Gamma' 
\ee
for some angle $\vartheta$. Since $\Gamma''$ is traceless and squares to the
identity, the `bound state' is another configuration preserving 1/2
supersymmetry. A trivial example is provided by the matrices
$\Gamma=\Gamma_{01}$ and $\Gamma' =\Gamma_{02}$ associated with M-waves in the
$1$-direction and $2$-direction, respectively. In this case $\Gamma''$
is clearly associated with a wave in some intermediate direction. To call this
a `bound state' is clearly an abuse of terminology (hence the quotes) but
the term can be understood in its usual sense in less trivial cases. An example
is provided by
\be\label{foure} 
\Gamma = \Gamma_{012}\, , \qquad \Gamma' = \Gamma_{012345}\, .
\ee
In this case, the matrix $\Gamma''$ can be associated with a bound
state of an M-5-brane with an M-2-brane.  From the perspective of the M-5-brane's
effective worldvolume field theory, the M-2-brane charge is the magnitude of the
topological 2-form charge 
\be\label{fourf}
Z^{MN} = \int\! dX^M\wedge dX^N \wedge H
\ee
where $H=dU$ is the 3-form field strength and the integral is over the
M-5-brane's `worldspace'.

Given that there is an M-5-brane/M-2-brane bound state, the fact that its
binding energy is strictly positive suggests that there is an attractive force
between a M-2-brane and an M-5-brane when one is parallel to the other
and separated by some distance. The M-2-brane would then be attracted to the
core of the M-5-brane, thereby lowering the energy until the energy bound is
saturated. In this case, one would not expect to find the relative separation
of the branes entering as a free parameter in the supergravity solution
representing the long-range fields of the M-5-brane/M-2-brane bound state. 
On the other hand, the 1/2 supersymmetry leads one to expect that it should be
possible to superpose parallel M-5-branes carrying the same M-2-brane charge. 
These considerations, together with the fact that the solution must reduce to 
that of the M-5-brane for zero M-2-brane charge, suggest that the supergravity
solution will again depend on a single harmonic function. There is indeed a
D=11 supergravity solution with these properties. The 11-metric of this
solution is \cite{ILPT}
\bea\label{boundstate}
ds^2_{11} &=& H^{1/3} \big( \sin^2\vartheta + H\cos^2\vartheta\big)^{1/3} 
\bigg[ H^{-1}ds^2(\bE^{(1,2)}) + \nn
&&+ \big( \sin^2\vartheta + H\cos^2\vartheta\big)^{-1} ds^2(\bE^3) +
ds^2(\bE^5)\bigg]
\eea
where $H$ (not to be confused with the worldvolume 3-form field strength of the
M-5-brane) is a harmonic function on $\bE^5$. As expected, it depends on the
angle $\vartheta$ which is arbitrary, classically, but is restricted in the
quantum theory because of the DNT quantization condition
satisfied by the M-2-brane and M-5-brane charges. This metric interpolates
between that of the M-2-brane and that of the M-5-brane; the same is true of the
complete solution.

Given a 1/2-supersymmetric M-2-brane/M-5-brane bound state various others
follow by duality. Compactifying to D=10 on one of the M-2-brane directions we
arrive at the IIA configuration represented by the array
$$
\begin{array}{lcccccccccc}
D4: & 1 & 2 & 3 & 4 & - & - & - & - & - \nn
F1: & 1 & - & - & - & - & - & - & - & -
\end{array}
$$
where `F1' stands for `Fundamental string'. T-dualizing in the 4-direction we
obtain a similar 1/2 supersymmetric bound state of a `Fundamental' IIB string
with a D-3-brane. This can be converted by S-duality into a bound state of a
D-string with a D-3-brane because the latter is S-self-dual. We now have the
IIB array
$$
\begin{array}{lcccccccccc}
D3: & 1 & 2 & 3 & - & - & - & - & - & - \nn
D1: & 1 & - & - & - & - & - & - & - & -
\end{array}
$$
T-duality in the 2 and 3 directions converts this into a bound state of a
D-string with a fundamental string, as required by the $Sl(2;Z)$ duality of IIB
superstring theory \cite{schwarz,Wittenb}. 

Let us now turn to the case in which $\Gamma$ and $\Gamma'$ commute. In this
case they may be simultaneously diagonalized with eigenvalues $\pm 1$. It
immediately follows that positivity implies the bound
$|\zeta| +|\zeta'| \le 1$, which is equivalent to a bound of the form
\be\label{fourg}
T \ge |Z| + |Z'|\, .
\ee
When this bound is saturated we can rewrite (\ref{foura}) as
\be\label{fourh}
\{Q,Q\} = 2P^0\big[ \zeta {\cal P} + \zeta' {\cal P}'\big] \, .
\ee
where ${\cal P} = (1/2)(1-\Gamma)$ and ${\cal P}' = (1/2)(1-\Gamma')$.
Since the projectors ${\cal P}$ and ${\cal P}'$ commute, a zero eigenvalue
eigenspinor of $\{Q,Q\}$ must be annihilated by both of them, i.e. it must
satisfy the joint conditions
\be\label{fouri}
\Gamma\epsilon =\epsilon \qquad \Gamma'\epsilon =\epsilon\, .
\ee 
Provided that the product $\Gamma\Gamma'$ is traceless (a condition that is
always met), the commuting matrices $\Gamma$ and $\Gamma'$ can be brought to the
diagonal form
\bea\label{fourj}
\Gamma &=& {\rm diag.} (\overbrace{1,\cdots ,1}^{16},
 \overbrace{-1,\cdots, -1}^{16}), \nn
\Gamma' &=& {\rm diag.} (\overbrace{1,\cdots ,1}^{8},
 \overbrace{-1,\cdots, -1}^{8}, \overbrace{1,\cdots ,1}^{8},
 \overbrace{-1,\cdots, -1}^{8}),
\eea
from which it is evident that the constraints (\ref{fouri}) preserve 1/4
supersymmetry. 

An example of a 1/4 supersymmetric configuration is provided by the orthogonal
intersection, on a point, of two M-2-branes, such that $\Gamma=\Gamma_{012}$ and
$\Gamma' =\Gamma_{034}$ \cite{PT}. This can be represented by the array
$$
\begin{array}{lccccccccccc}
M2: & 1 & 2 & - & - & - & - & - & - & - & -\nn
M2: & - & - & 3 & 4 & - & - & - & - & - & -
\end{array}
$$
Most other 1/4 supersymmetric configurations of orthogonally intersecting
branes may be obtained from this one by various `duality chains'. For example,
compactifying on the $\natural\;$th direction we obtain a similar configuration
of two intersecting D-2-branes. Then, T-dualizing in the 5 and 6 directions we
arrive a configuration of intersecting D-4-branes
$$
\begin{array}{lccccccccccc}
D4: & 1 & 2 & - & - & 5 & 6 & - & - & - \nn
D4: & - & - & 3 & 4 & 5 & 6 & - & - & - 
\end{array}
$$
which may lifted to D=11 to yield a configuration of two M-5-branes
intersecting on a 3-brane \cite{PT}
$$
\begin{array}{lccccccccccc}
M5: & 1 & 2 & - & - & 5 & 6 & - & - & - & \natural \nn
M5: & - & - & 3 & 4 & 5 & 6 & - & - & - & \natural
\end{array}
$$
In an alternative notation, the initial configuration of intersecting
M-2-branes is denoted by $(0|M2,M2)$ and the final configuration of
intersecting M-5-branes by $(3|M5,M5)$. The duality chain from one to the
other is then indicated as follows
\be\label{fourk}
(0|M2,M2) \ {\buildrel M \over \rightarrow} \ (0|D2,D2)\  
{\buildrel T^2 \over \rightarrow}\  (2|D4,D4) \ {\buildrel M \over \rightarrow}
\ (3|M5,M5)\, .
\ee

From the same starting point we can compactify instead on the 2-direction to
obtain a IIA string intersecting a D-2-brane. Relabelling the directions, we
have 
$$
\begin{array}{lccccccccccc}
F1: & 1 & - & - & - & - & - & - & - & - \nn
D2: & - & 2 & 3 & - & - & - & - & - & - 
\end{array}
$$
T-dualizing in the $4$ and $5$ directions we obtain an intersection of a
IIA string with a D-4-brane, which is a reduction to D=10 of 
$$
\begin{array}{lcccccccccccc}
M2: & 1 & - & - & - & - & - & - & - & - & \natural \nn
D2: & - & 2 & 3 & 4 & 5 & - & - & - & - & \natural
\end{array}
$$
i.e. the intersection on a string of an M-2-brane with an M-5-brane. This
sequence of steps can be denoted by the following duality chain
\be\label{fourl}
(0|M2,M2) \ {\buildrel M \over \rightarrow} \ (1|F1,D2)\  
{\buildrel T^2 \over \rightarrow} \ (1|F1,D4) \ {\buildrel M \over \rightarrow}
\ (1|M2,M5)\, .
\ee
The intersection of the M-2-brane with the M-5-brane can be interpreted as a
coincidence of two membrane boundaries, one of an `incoming' M-2-brane and the
other of an `outgoing' M-2-brane. This interpretation is possible because, as
mentioned previously, the M-2-brane can end on an M-5-brane. Given this, it 
then follows that the array representing the intersection of the IIA string with
the D-2-brane, for example, can be reinterpreted as the coincidence of two
endpoints of IIA strings. More generally, all `brane-boundary' possibilities,
including Type II strings ending on D-branes, follow by duality from the
possibility of an M-2-brane ending on an M-5-brane, so we may consider this to
be the key case to understand. It does not follow from the D=11 superalgebra
alone, but it will be deduced below from considerations related to the
worldvolume supersymmetry algebra of the M-5-brane.  

For terminological convenience we shall regard brane boundaries as
special cases of intersections. It should also be stated that `intersections'
include cases in which there is merely an `overlap'. This point may be
illustrated by the case of two `intersecting' M-2-branes. There is  a
genuine intersection only if the positions of the two M-2-branes in the 
`overall' transverse 6-space coincide; otherwise they might be said to be
`overlapping'. However, the fact that the configuration saturates an energy bound
of the form (\ref{fourg}) implies the absence of a force between the two
M-2-branes and hence that the distance $L$ of separation between them in the
overall transverse 6-space is a free parameter. In particular, we may
choose $L=0$, so a genuine intersection is included as a special case. This is
equally true for any of the 1/4 supersymmetric configurations represented by
the above arrays.

A further implication of the `no force' condition is that one can obtain a
supergravity solution representing a 1/4 supersymmetric intersecting brane
configuration by a type of superposition, which is summarized by the `harmonic
function rule' \cite{tseytlin,GKT}. The above M-brane intersections will serve to
illustrate the rule. We shall only consider its application to the 11-metric. In
the
$(0|M2,M2)$ case this is
\be\label{fourm}
ds^2_{11}  = (H_1H_2)^{1/3}\bigg[(H_1H_2)^{-1}dt^2 + H_1^{-1} |dz_1|^2 +  
H_2^{-1} |dz_2|^2 + ds^2(\bE^6) \bigg]
\ee
where $z_1$ and $z_2$ are  complex coordinates parametrizing the two
orthogonal 2-planes occupied by the two membranes. This 11-metric is essentially
determined by the requirement that if either $(H_1-1)$ or $(H_2-1)$ vanishes we
recover the 11-metric of a single M-2-brane. The other fields are determined by
the same requirement. The fact that the solution depends on {\sl two} harmonic
functions is directly related to the no-force condition implied by 1/4
supersymmetry. The only subtlety lies in the fact that to solve the field
equations of D=11 supergravity we must restrict the variables on which the
harmonic functions
$H_1$ and $H_2$ depend; for the moment we postpone discussion of this point. 

A feature of the 11-metric (\ref{fourm}) (shared by the 11-metric
(\ref{boundstate}) representing a 1/2 supersymmetric non-marginal bound state)
is that, apart from an overall conformal factor, there is a factor of $H_i^{-1}$
for each term in the metric corresponding to a direction `occupied' by the i'th
brane. Applying these principles to the
$(3|M5,M5)$ case we find that
\bea\label{fourn}
ds^2_{11} &=& (H_1H_2)^{2/3}\bigg[(H_1H_2)^{-1}ds^2(\bE^{(1,3)}) + H_1^{-1}
|dz_1|^2\nn
&& \qquad + \,  H_2^{-1} |dz_2|^2 + ds^2(\bE^3) \bigg]
\eea
where the complex variables $z_1$ and $z_2$ now parametrize the two
`relative transverse' spaces, i.e. the 2-space in each M-5-brane transverse
to the 3-brane intersection. A similar application to the $(1|M2,M5)$ case 
yields
\bea\label{fouro}
ds^2_{11} &=& H_1^{1/3}H_2^{2/3}\bigg[(H_1H_2)^{-1}ds^2(\bE^{(1,1)}) + H_1^{-1}
dx^2 \nn
&& \qquad +\, H_2^{-1}|dq|^2 + ds^2(\bE^4) \bigg]
\eea
where $x$ is a (real) coordinate parametrizing the relative transverse space of
the membrane and $q$ is a quaternonic coordinate parametrizing the relative
transverse space of the M-5-brane. In all of these cases, the harmonic
functions must be taken to be independent of all but the overall transverse
coordinates. In other words, the D=11 supergravity solutions of this form are
necessarily translational invariant in all directions tangent to a
participating brane. This is an appropriate restriction if we wish to
consider toroidal compactification in which each p-brane is wrapped on a
p-cycle, since we may then essentially read off the solution for an extreme
black hole of the compactified supergravity theory. Otherwise the restriction
is not appropriate but it seems to be the best that can be achieved by
superposition; it is not excluded that there exists a more general class of
solution but if so it will not be found by the harmonic function
rule\footnote{We refer to \cite{gaunt} for a review of recent progress in this
direction.}.

Given two commuting matrices $\Gamma$ and $\Gamma'$ associated 
with a 1/4 supersymmetric configuration of two intersecting branes the product
$\Gamma\Gamma'$ is also traceless and squares to the identity. It
also commutes with both $\Gamma$ and $\Gamma'$. Thus it is always possible to
add an additional brane `for free' in the sense that the configuration still
preserves 1/4 supersymmetry. The 1/4 supersymmetric configuration of two
orthogonally intersecting M2-branes will illustrate the point. Since both
$\Gamma_{012}\epsilon=\epsilon$ and $\Gamma_{034}\epsilon=\epsilon$ it is 
automatic that $\Gamma_{1234}\epsilon=-\epsilon$. In this case the `third 
brane' is actually an M-KK-monopole, or rather an anti M-KK-monopole; note that 
if we wish to maintain 1/4 supersymmetry then we are not free to choose the
orientation of the `third brane'. The 1/4 supersymmetric configuration of two
M-2-branes and an M-KK-monopole may be represented by the array
$$
\begin{array}{lccccccccccc}
M2: & 1 & 2 & - & - & - & - & - & - & - & - \nn
M2: & - & - & 3 & 4 & - & - & - & - & - & - \nn
MKK: & \times  & - & - & - & 5 & 6 & 7 & 8 & 9 & \natural
\end{array}
$$
where $\times$ indicates the KK circle. Reducing to D=10 in this direction and
relabelling the result yields
$$
\begin{array}{lccccccccc}
F1: & 1 & - & - & - & - & - & - & - & - \nn
D2: & - & 2 & 3 & - & - & - & - & - & - \nn
D6: & - & - & - & 4 & 5 & 6 & 7 & 8 & 9 
\end{array}
$$
T-dualizing in the $4$ and $5$ directions now yields
$$
\begin{array}{lccccccccc}
F1: & 1 & - & - & - & - & - & - & - & - \nn
D4: & - & 2 & 3 & 4 & 5 & - & - & - & - \nn
D4: & - & - & - & - & - & 6 & 7 & 8 & 9 
\end{array}
$$ 
which may be lifted to D=11 as 
$$
\begin{array}{lcccccccccc}
M2: & 1 & - & - & - & - & - & - & - & - & \natural \nn
M5: & - & 2 & 3 & 4 & 5 & - & - & - & - & \natural \nn
M5: & - & - & - & - & - & 6 & 7 & 8 & 9 & \natural
\end{array}
$$
The M-2-brane now intersects each of two M-5-branes on a string.
Alternatively, we may view the configuration as one in which a membrane is
`stretched' between two M-5-branes, in such a way that it has a string
boundary on each. If we remove the membrane we obtain a new 1/4 supersymmetric
configuration, $(1|M5,M5)$, in which two M-5-branes `intersect' on a string
although, as before, there is a genuine intersection only when the separation 
in the $1$ direction vanishes. The harmonic function rule applied to the
$(1|M5,M5)$ configuration yields
\be\label{fourp}
ds^2_{11} = (H_1H_2)^{2/3} \bigg[ (H_1H_2)^{-1}ds^2(\bE^{(1,1)}) + H_1^{-1}
|dq_1|^2 + H_2^{-1}|dq_2|^2 + dx^2\bigg]
\ee
where $q_1$ and $q_2$ are quaternionic coordinates parametrizing the two
relative transverse spaces to the string intersection in the M-5-branes, and
$x$ is the coordinate in which the two M-5-branes are, in principle,
separated. In fact, for this to be a solution of D=11 supergravity both harmonic
functions must be taken to be {\sl independent of the overall transverse
dimension} $x$; instead, $H_1$ is harmonic in $q_2$ and $H_2$ is harmonic in
$q_1$ \cite{GKT}. Again, there may be a more general solution in which the
M-5-branes are localized in the $x$ direction but, if so, it is not given by the
harmonic function rule.

When the branes of a 1/4 supersymmetric `intersecting brane' configuration
actually intersect, the intersection should appear in the worldvolume field
theory of each as a 1/2 supersymmetric worldvolume `soliton' of some kind. 
This is one way of understanding how 1/4 supersymmetric intersecting brane
configurations arise\footnote{At this point, the content of these lectures 
diverges from the content of those given in Carg{\`e}se.}. A simple example of
this point of view is provided by consideration of the `intersecting brane'
configuration
$(0|D0,D4)$ obtained from $(0|M2,M2)$ by the duality chain 
\be\label{afourm}
(0|M2,M2) \ {\buildrel M \over \rightarrow}\  (0|D2,D2) \
{\buildrel T^2 \over \rightarrow} (0|D0,D4) \, .
\ee
The worldvolume field theory for a single D-4-brane is a (4+1)-dimensional
maximally supersymmetric Dirac-Born-Infeld theory. In the IIA Minkowski vacuum,
and assuming a Minkowski worldvolume (i.e. choosing the `static' gauge and
setting to zero all worldvolume scalars describing transverse fluctuations),
the bosonic Lagrangian reduces to the pure Born-Infeld form 
\be\label{fourma}
{\cal L}= -\sqrt{-\det \big(\eta + F\big)}
\ee
where $\eta$ is the D=5 Minkowski metric and $F$ is the BI 2-form. 
We now set the electric components of $F$ to zero, in which case we can
interpret ${\cal L}$ as minus the energy density ${\cal E}$. The $5\times 5$
determinant also reduces to minus the $4\times 4$ determinant of the matrix
$(1+F)$, where $F$ is now the magnetic 2-form component of the BI field
strength. Thus ($a,b=1,2,3,4$)
\bea\label{fourmb}
{\cal E}^2 &=& \det(\delta_{ab} + F_{ab}) \nn
&=& (1 \pm{1\over4} \tr\; F\tilde F)^2 - {1\over4}\tr (F\pm\tilde F) ^2
\eea
where $\tilde F$ is the worldspace Hodge dual of $F$. The trace is over
the worldspace indices, i.e. $\tr F ^2 = F_{ab}F^{ba} \le 0$, 
but the notation can be extended to include a trace over
$su(n)$ indices, as appropriate to the $U(n)$ gauge theory on $n$ coincident
D-4-branes. In either case, we deduce that \cite{GGT}
\be\label{fourmc}
{\cal E} \ge 1 \pm {1\over4} \tr F\tilde F 
\ee
with equality when $F=\tilde F$. The total energy $E$, relative to the 
worldvolume vacuum, is just the integral of ${\cal E}-1$ over the worldspace.
Thus we derive the bound 
\be\label{fourmd}
E \ge |Z|
\ee
where $Z$ is the topological charge
\be\label{fourme}
Z= {1\over4}\tr \int \! F\tilde F \, .
\ee
The bound is saturated by solutions of $F=\tilde F$. These are
just multi-instantons in the non-abelian case, which is appropriate to coincident
D-4-branes. In the abelian case any solution must have a singular $U(1)$ gauge
potential, although the energy will remain finite for finite charge. Since not
much is known about abelian BI instantons, let us concentrate on the non-abelian
case. The soliton energy is independent of the instanton size, which is 
therefore a modulus of the solution. There is nothing to prevent us from
shrinking the instanton to zero size. In fact, if we were to allow  a slow time
variation of the instanton parameters then the instanton could shrink to zero
size in {\sl finite} time. This is a reflection of the fact that the natural
metric on instanton moduli space is geodesically incomplete, indicating that new
physics is needed to determine what happens when an instanton shrinks to zero
size. The new physics is provided by the interpretation of the (4+1)-dimensional
Minkowski vacuum as the worldvolume of a D-4-brane: when the instanton shrinks
to zero size it simply leaves the worldvolume as a D-0-brane \cite{douglas}. 

The existence of a 1/2 supersymmetric soliton in the worldvolume field theory
of the D-4-brane, or multi D-4-brane, is suggested by the presence of a central
charge in its five-dimensional worldvolume supersymmetry algebra. This charge has
a natural interpretation as the momentum in an additional space direction. In
fact, the algebra is the reduction of the D=6 (2,0) supersymmetry algebra, and
the (scalar) central charge arising in this reduction is precisely the
6-component of the 6-momentum. This is as expected because the D-4-brane has an
M-theory interpretation as an $S^1$-wrapped M-5-brane, for which the
(gauge-fixed) worldvolume field theory has D=6 (2,0) supersymmetry. Thus,
M-theory predicts the existence of a tower of KK quantum states in the D-4-brane
worldvolume field theory \cite{DKPS,PTtwo} with one such state for each integer
value of the soliton charge, alias instanton number. This amounts to a
prediction of a marginal bound state  in the system of $n$ D-0-branes and a
D-4-brane for each $n$. Any such bound state must indeed be marginal, and there
is good evidence that the prediction is correct \cite{DKPS}. 

We have just invoked the presence of a central charge in the worldvolume
supersymmetry algebra of the D-4-brane to explain the presence of a marginal
1/2 supersymmetric bound state with a D-0-brane, for which the long range
supergravity fields are those of a 1/4 supersymmetric $(0|D0,D4)$ `intersecting
brane' configuration. However, as by now should be clear, there is no reason to
concentrate exclusively on {\sl scalar} charges; we should consider all
charges. Since the D-4-brane is a wrapped M-5-brane it is more economical to
consider the p-form charges in the M-5-brane's worldvolume supersymmetry
algebra. Allowing for all possible p-form charges, we have \cite{HLW2}
\be\label{afourn}
\{Q_\a^I,Q_b^J\} = \Omega^{IJ} P_{[\a\b]} + Y^{[IJ]}_{[\a\b]} +
Z^{(IJ)}_{(\a\b)}
\ee
where $\a,\b =1,\dots,4$ is a spinor index of the Lorentz group $SU^*(4)\cong
Spin(5,1)$ and $I=1,\dots,4$ is an index of the internal `R-symmetry' group
$Sp(2)$, with $\Omega^{IJ}$ being its invariant antisymmetric tensor. The
spinor supercharge therefore has 16 complex components but is subject to a
`symplectic Majorana' condition that reduces the number of independent
components by a factor of 2. The $Y$-charge is a worldvolume 1-form and the
$Z$-charge a worldvolume self-dual 3-form. For simplicity, we shall consider
here only the space components of these charges, which would be carried by 
worldvolume $p$-branes for $p=1$ and $p=3$, respectively. The existence of these
worldvolume branes could be anticipated from the fact that there exist 1/4
supersymmetric intersecting brane configurations of M-theory in which an
M-5-brane intersects another M-brane on a string (the $(1|M2,M5)$ configuration)
or a 3-brane (the $(3|M5,M5)$ configuration). 

It might appear that there are
more $p$-brane charges than are required because the $Sp(2)$ representations of
these charges are
\be 
Y (p=1) :  {\bf 5} \qquad Z (p=3) : {\bf 10}\, .
\ee
However, $Sp(2)$ is isomorphic to $Spin(5)$, which we may interpret as the
rotation group in the 5-space transverse to the M--brane worldvolume in the
D=11 spacetime. In this case, the $Sp(2)$ representations simply provide the
information needed for the {\sl spacetime} interpretation of the worldvolume
branes as intersections with other branes. Specifically, the ${\bf 5}$ and 
${\bf 10}$ representations can be interpreted as, respectively, a 1-form and
a 2-form in the transverse 5-space. A p-form charge that is a transverse q-form
is naturally interpreted as the charge of a (p+q)-brane in spacetime with a
p-brane intersection with (in this case) the M-5-brane. We thereby recover the
from the worldvolume supersymmetry algebra the spacetime interpretation of the
1-brane and 3-brane on the M-5-brane that we earlier deduced by consideration of
the spacetime supersymmetry algebra \cite{BGT}.

Just as the scalar charge $Z$ in the D-4-brane's worldvolume superalgebra
was expressible as the topological charge integral (\ref{fourme}), so
the worldvolume charges $Y$ and $Z$ in the M-5-brane's worldvolume
superalgebra must be expressible as integrals of worldvolume fields over the
transverse subspace of worldspace. Let us concentrate on the string charge.
Its magnitude must be expressible as an integral over the 4-dimensional subspace
of worldspace transverse to the string; let us call this $w_4$. Since $Y$ is a
5-vector in the transverse subspace of spacetime, it must depend on one of the
five worldvolume scalar fields ${\bf X}$ describing fluctuations of the 
M-5-brane in these directions; the choice of scalar corresponds to a choice of
direction of the 5-vector. These considerations imply that $Y$ must take the
form \cite{GGT} 
\be\label{topcharge}
Y = \int_{w_4} H \wedge dX
\ee
where $H$ is the closed 3-form field strength of the 2-form worldvolume gauge
potential and $X$ is the scalar. 

If we set to zero all other physical scalars then, in the static gauge, a static
configuration is one for which
\be
X^M= (t,\sigma^a,X(\sigma),0,\dots,0)
\ee
where $(t,\sigma^a)$ are the worldvolume coordinates. The conjugate momentum is
\be
P_M = (-{\cal E}, V_a, 0,0,\dots,0)
\ee
where ${\cal E}$ is the energy density, $V_a$ is the worldspace vector density
introduced in  (\ref{phaseb}), and we have used the fact that the diffeomorphism
constraint reduces in static gauge, and for static configurations, to $P_a=V_a$.
Under  these conditions the hamiltonian constraint of (\ref{phaseb}) implies that
\be
{\cal E}^2 = 1 + (\partial X)^2 + {1\over2} |\tilde {\cal H}|^2 + 
|\tilde {\cal H}\cdot \partial X|^2  + |V|^2
\ee
where the vertical bars indicate contraction with the kronecker delta. Here we
are following \cite{GGT}, in which it is pointed out that this expression for
the energy density of static worldvolume field configurations can be
rewritten as
\be
{\cal E}^2 = \big|\zeta^a \pm \tilde {\cal H}^{ab}\partial_b X\big| ^2 
+ 2\bigg|\partial_{[a} X \zeta_{b]} \mp {1\over2} \delta_{ac}\delta_{bd} \tilde
{\cal H}^{cd}\bigg|^2  + {1\over2} (\zeta^a\partial_a X)^2
\ee
where $\zeta$ is a constant unit 5-vector. One can deduce from this (e.g. by
making a particular choice for $\zeta$) that 
\be
{\cal E} -1 \ge \pm i_\zeta (\star H dX)
\ee
where $\star$ is the Hodge dual on worldspace and $i_\zeta$ indicates
contraction with the (constant) vector field $\zeta$. The inequality is
saturated when
\be\label{stringeq}
H =\pm i_\zeta (\star dX) \qquad {\cal L}_\zeta X=0
\ee
where ${\cal L}_\zeta$ is the Lie derivative with respect to $\zeta$. Note that
these conditions imply that ${\cal L}_\zeta H=0$ (and hence that $V_a=0$). Since
the field configuration saturating the bound is translational invariant in the
$\zeta$ direction we should integrate the energy density over the subspace $w_4$
orthogonal to this direction and interpret the result as a string tension $T$.
The result is that 
\be
T \ge |Y|
\ee
where $Y$ is the topological charge of (\ref{topcharge}), and the bound is
saturated when 
\be\label{bogeq}
H =\pm * dX
\ee
where it is now understood that all fields are defined on $w_4$ and $*$ is the
Hodge dual in this space.

Since $dH=0$, we learn from (\ref{bogeq}) that $X$ is harmonic on $w_4$.
Isolated singularities of this harmonic function represent infinite straight
string defects in the M-5-brane \cite{HLW1}. We may assume that $X$ vanishes at
spatial infinity, so that if the only singularity is at the origin then
$X=X(r)$ where
\be
X(r) = {1\over r^2}\; .
\ee
As we approach the singularity at $r=0$, the position of the M-5-brane in the
transverse 5-space extends further and further along the X-axis, thereby by
creating an infinite `spike' in this direction\footnote{This feature of
1/2 supersymmetric worldvolume solitons was first noted in the context of
D-branes \cite{CM,GWG} and extended to the string soliton in the
M-5-brane in \cite{HLW1}. The presentation here follows \cite{GGT}.}.
The `spike' is actually a `ridge' due to the translational invariance in the
$\zeta$-direction. To determine the nature of this ridge we note that the
topological charge
$Y$ may be rewritten as the small radius limit of a surface integral over a
3-sphere centred on the origin. Since $X$ is constant over this integration
surface we then have 
\be
Y = \big[\lim_{\delta \rightarrow 0} X(\delta)\big] \int_{S^3} H\, .
\ee
The integral of $H$ is what one might naively have taken to be the string
charge. We see that the charge is instead the product of the naive charge with 
an infinite factor, since $X(\delta)$ diverges as $\delta\rightarrow 0$. 
However, this divergence is exactly what we would expect if the string is the
boundary of a membrane of constant surface tension $T_2= \int_{S^3} H$. Thus, 
the `ridge' of the solution is actually a membrane with its boundary on the
M-5-brane. Remarkably, the worldvolume string soliton of the
M-5-brane provides its own {\sl spacetime} interpretation as the
boundary of an M-2-brane. 

\section{Epilogue}

We have seen that an enormous amount of information about M-theory and its
duality relations to superstring theories is encoded in the M-theory and Type
II supersymmetry algebras.  We have also begun to see how 1/4-supersymmetric
intersections of the `basic', 1/2-supersymmetric, branes are encoded in their
worldvolume supersymmetry algebras. The next step would be to consider
supersymmetry fractions less than 1/4; for example 1/8, which corresponds to
N=1 in D=4. This and other fractions are realized by certain non-orthogonal
intersections and some triple intersections. Having classified possible
fractions of supersymmetry associated with intersecting brane configurations
our next step would be to determine the supersymmetric field theories on the
intersections. It seems possible that all supersymmetric field theories may be
obtainable this way.  However, there is much less that can be said about these
further developments from the algebraic viewpoint advocated here, at least for
the present, so this is a good place to end these lectures. 

\vskip 0.5cm
\noindent
{\bf Acknowledgements}: The written version of these lectures was completed
while I was on leave at the University of Barcelona. I thank the faculty of
Physics for their hospitality, and I am grateful for the support of the
Iberdrola  {\it Profesor Visitante} program. In addition, I thank my recent
collaborators, Eric Bergshoeff, Martin Cederwall, Philip Cowdall, Jerome
Gauntlett, Gary Gibbons, Joaquim Gomis, Chris Hull, George Papadopoulos and
Dmitri Sorokin for conversations related to work reviewed here, and Dan
Freedman for pointing out some errors in an earlier version.
\vfill\eject

\end{document}